\documentclass[9pt,shortpaper,twoside,web]{ieeecolor}

\usepackage{generic}
\usepackage{cite}
\usepackage{amsmath,amssymb,amsfonts}
\usepackage{graphicx}
\usepackage{subcaption}  
\usepackage{textcomp}
\usepackage{pifont}

\def\BibTeX{{\rm B\kern-.05em{\sc i\kern-.025em b}\kern-.08em
    T\kern-.1667em\lower.7ex\hbox{E}\kern-.125emX}}
\usepackage{multirow}
\usepackage{comment}
\usepackage{verbatim} 

\usepackage{tikz}
\usepackage{pgfplots}
\usepackage{pgfplotstable}
\usepackage{graphicx}
\usepackage{pifont}
\usepackage[export]{adjustbox}
\usepackage{pgf-pie}
\usepackage{tkz-graph}
\usepackage{graphicx}
\usepackage{subcaption}
\usepackage[skip=2pt]{caption}
\usepackage{hyperref}
\usepackage{xcolor}
\usepackage{tcolorbox}
\usepackage{pifont}
\usepackage{listings}
\hypersetup{
    colorlinks,
    linkcolor={red!40!black},
    citecolor={blue!40!black},
    urlcolor={black}
}
\usepackage{algorithm}
\usepackage{algpseudocode}
\usepackage{amsmath}

\usepackage{booktabs}
\usepackage{titlesec}
\usepackage{placeins}
\usepackage{float}
\usepackage{dblfloatfix}
\usepackage{makecell}

\usetikzlibrary{pgfplots.groupplots}
\usetikzlibrary{shadows,patterns,shapes,arrows,decorations.pathmorphing,backgrounds,positioning,fit,plotmarks,calc,spy,matrix}
\pgfplotsset{compat=1.11,
    /pgfplots/ybar legend/.style={
    /pgfplots/legend image code/.code={%
       \draw[##1,/tikz/.cd,yshift=-0.25em]
        (0cm,0cm) rectangle (3pt,0.8em);},
   },
}
\usepackage{titlesec}
\usetikzlibrary{decorations.text}
\definecolor{boxbg}{gray}{0.95} 
\definecolor{cadetblue}{rgb}{0.37, 0.62, 0.63}
\definecolor{airforceblue}{rgb}{0.36, 0.54, 0.66}
\definecolor{caribbeangreen}{rgb}{0.0, 0.8, 0.6}
\definecolor{carolinablue}{rgb}{0.6, 0.73, 0.89}
\definecolor{darkgoldenrod}{rgb}{0.72, 0.53, 0.04}
\definecolor{debianred}{rgb}{0.84, 0.04, 0.33}
\definecolor{fuzzywuzzy}{rgb}{0.8, 0.4, 0.4}
\definecolor{grullo}{rgb}{0.66, 0.6, 0.53}
\definecolor{ceil}{rgb}{0.57, 0.63, 0.81}
\definecolor{candypink}{rgb}{0.89, 0.44, 0.48}
\definecolor{calpolypomonagreen}{rgb}{0.12, 0.3, 0.17}
\definecolor{burntsienna}{rgb}{0.91, 0.45, 0.32}
\definecolor{atomictangerine}{rgb}{1.0, 0.6, 0.4}
\definecolor{goldenrod}{rgb}{0.85, 0.65, 0.13}
\definecolor{gamboge}{rgb}{0.89, 0.61, 0.06}
\definecolor{amber}{rgb}{1.0, 0.75, 0.0}
\definecolor{battleshipgrey}{rgb}{0.52, 0.52, 0.51}
\definecolor{darkcerulean}{rgb}{0.03, 0.27, 0.49}
\definecolor{fuzzywuzzy}{rgb}{0.8, 0.4, 0.4}
\definecolor{mediumseagreen}{rgb}{0.24, 0.7, 0.44}
\definecolor{antiquebrass}{rgb}{0.8, 0.58, 0.46}
\definecolor{apricot}{rgb}{0.98, 0.81, 0.69}
\definecolor{asparagus}{rgb}{0.53, 0.66, 0.42}
\definecolor{bananamania}{rgb}{0.98, 0.91, 0.71}
\definecolor{cadmiumgreen}{rgb}{0.0, 0.42, 0.24}
\definecolor{chocolate}{rgb}{0.48, 0.25, 0.0}
\definecolor{cinereous}{rgb}{0.6, 0.51, 0.48}
\definecolor{aliceblue}{rgb}{0.94, 0.97, 1.0}
\definecolor{beaublue}{rgb}{0.74, 0.83, 0.9}
\definecolor{blizzardblue}{rgb}{0.67, 0.9, 0.93}
\definecolor{bittersweet}{rgb}{1.0, 0.44, 0.37}
\definecolor{camouflagegreen}{rgb}{0.47, 0.53, 0.42}
\definecolor{darkolivegreen}{rgb}{0.33, 0.42, 0.18}
\definecolor{darkpastelblue}{rgb}{0.47, 0.62, 0.8}
\definecolor{desertsand}{rgb}{0.93, 0.79, 0.69}
\definecolor{deeppeach}{rgb}{1.0, 0.8, 0.64}
\definecolor{indianred}{rgb}{0.8, 0.36, 0.36}
\definecolor{oldmauve}{rgb}{0.4, 0.19, 0.28}
\definecolor{lightblue}{rgb}{0.68, 0.85, 0.9}
\definecolor{lightcyan}{rgb}{0.88, 1.0, 1.0}
\definecolor{viridian}{rgb}{0.25, 0.51, 0.43}
\definecolor{slategray}{rgb}{0.44, 0.5, 0.56}
\definecolor{manatee}{rgb}{0.59, 0.6, 0.67}
\definecolor{darkbrown}{rgb}{0.4, 0.26, 0.13}
\definecolor{almond}{rgb}{0.94, 0.87, 0.8}
\def\BibTeX{{\rm B\kern-.05em{\sc i\kern-.025em b}\kern-.08em
    T\kern-.1667em\lower.7ex\hbox{E}\kern-.125emX}}
\usepackage{colortbl}
\usepackage{tabulary}
\pgfkeys{%
/piechartthreed/.cd,
scale/.code                =  {\def\piechartthreedscale{#1}},
mix color/.code            =  {\def\piechartthreedmixcolor{#1}},
background color/.code     =  {\def\piechartthreedbackcolor{#1}},
name/.code                 =  {\def\piechartthreedname{#1}}}
\newcommand\piechartthreed[2][]{%
   \pgfkeys{/piechartthreed/.cd,
     scale            = 1,
     mix color        = gray,
     background color = white,
     name             = pc} 
  \pgfqkeys{/piechartthreed}{#1}
  \begin{scope}[scale=\piechartthreedscale] 
  \begin{scope}[xscale=5,yscale=3] 
     \path[preaction={fill=black,opacity=.8,
         path fading=circle with fuzzy edge 20 percent,
         transform canvas={yshift=-15mm*\piechartthreedscale}}] (0,0) circle (1cm);
     \pgfmathsetmacro\totan{0} 
     \global\let\totan\totan 
     \pgfmathsetmacro\bottoman{180} \global\let\bottoman\bottoman 
     \pgfmathsetmacro\toptoman{0}   \global\let\toptoman\toptoman 
     \begin{scope}[draw=black,thin]
     \foreach \an/\col [count=\xi] in {#2}{%
     \def\space{ } 
        \coordinate (\piechartthreedname\space\xi) at (\totan+\an/2:0.75cm); 
        \ifdim 180pt>\totan pt 
         \ifdim 0pt=\toptoman pt
            \pgfmathsetmacro\toptoman{180} 
            \global\let\toptoman\toptoman         
            \else
          \fi
        \fi   
        \fill[\col!80!gray,draw=black] (0,0)--(\totan:1cm)  arc(\totan:\totan+\an:1cm)
                                     --cycle;     
       \pgfmathsetmacro\finan{\totan+\an}
       \ifdim 180pt<\finan pt 
         \ifdim 180pt=\bottoman pt
            \shadedraw[left color=\col!20!\piechartthreedmixcolor,
                       right color=\col!5!\piechartthreedmixcolor,
                       draw=black,very thin] (180:1cm) -- ++(0,-3mm) arc (180:\totan+\an:1cm) 
                                                       -- ++(0,3mm)  arc (\totan+\an:180:1cm);
            \pgfmathsetmacro\bottoman{0}
            \global\let\bottoman\bottoman
            \else
            \shadedraw[left color=\col!20!\piechartthreedmixcolor,
                       right color=\col!5!\piechartthreedmixcolor,
                       draw=black,very thin](\totan:1cm)-- ++(0,-3mm) arc(\totan:\totan+\an:1cm)
                                                        -- ++(0,3mm)  arc(\totan+\an:\totan:1cm); 
          \fi
        \fi
        \pgfmathsetmacro\totan{\totan+\an}  \global\let\totan\totan 
       } 
    \end{scope}
   \end{scope}  
\end{scope}
}
\newcommand{\cmark}{\ding{52}}%

\usepackage{etoolbox}
\usepackage{xcolor}
\usepackage{pdfpages} 
\usepackage{listings}

\usepackage{colortbl}
\usepackage{tcolorbox}
\usepackage{booktabs}
\usepackage{pgfplots}
\usepackage{tikz}
\usepackage{pgf-pie}
\usepackage{url}

\usepackage[colorinlistoftodos, textsize=tiny]{todonotes} 
\usepackage[normalem]{ulem} 

\usepackage{ifthen}
\newif\ifremark
\remarkfalse

\definecolor{dkgreen}{rgb}{0,0.6,0}
\definecolor{gray}{rgb}{0.5,0.5,0.5}
\definecolor{mauve}{rgb}{0.58,0,0.82}
\definecolor{OliveGreen}{rgb}{0,0.6,0}
\definecolor{mahogany}{rgb}{0.75, 0.25, 0.0}
\definecolor{darkmidnightblue}{rgb}{0.0, 0.2, 0.4}
\definecolor{blue}{rgb}{0.0, 0.0, 1.0}
\definecolor{navyblue}{rgb}{0.0, 0.0, 0.5}
\definecolor{apricot}{rgb}{0.98, 0.81, 0.69}
\definecolor{antiquewhite}{rgb}{0.98, 0.92, 0.84}
\definecolor{brickred}{rgb}{0.8, 0.25, 0.33}
\definecolor{bananamania}{rgb}{0.98, 0.91, 0.71}
\definecolor{bisque}{rgb}{1.0, 0.89, 0.77}
\definecolor{champagne}{rgb}{0.97, 0.91, 0.81}
\definecolor{eggshell}{rgb}{0.94, 0.92, 0.84}
\definecolor{darkolivegreen}{rgb}{0.33, 0.42, 0.18}
\definecolor{phthalogreen}{rgb}{0.07, 0.21, 0.14}
\definecolor{richblack}{rgb}{0.0, 0.25, 0.25}
\definecolor{anti-flashwhite}{rgb}{0.95, 0.95, 0.96}
\definecolor{burntOrange}{HTML}{b35806}  
\definecolor{softOrange}{HTML}{f1a340}   
\definecolor{lightPeach}{HTML}{fee0b6}   
\definecolor{lavender}{HTML}{998ec3}     

\definecolor{pie1}{gray}{0.2}
\definecolor{pie2}{gray}{0.4}
\definecolor{pie3}{gray}{0.6}
\definecolor{pie4}{gray}{0.8}
\definecolor{pie5}{gray}{0.9}

\definecolor{pie11}{gray}{0.1}
\definecolor{pie12}{gray}{0.25}
\definecolor{pie13}{gray}{0.4}
\definecolor{pie14}{gray}{0.55}
\definecolor{pie15}{gray}{0.70}
\definecolor{pie16}{gray}{0.85}

\definecolor{pastel11}{HTML}{FFEEF4}
\definecolor{pastel12}{HTML}{E4E4D0}
\definecolor{pastel13}{HTML}{AEC3AE}
\definecolor{pastel14}{HTML}{94A684}
\definecolor{pastelRed}{HTML}{FF90BC}

\definecolor{pastel41}{HTML}{A1CCD1}
\definecolor{pastel42}{HTML}{F4F2DE}
\definecolor{pastel43}{HTML}{E9B384}
\definecolor{pastel44}{HTML}{7C9D96}

\definecolor{color11}{HTML}{fee8c8}
\definecolor{color12}{HTML}{fdbb84}
\definecolor{color13}{HTML}{e34a33}

\definecolor{pastel31}{HTML}{f1eef6}
\definecolor{pastel32}{HTML}{d7b5d8}
\definecolor{pastel33}{HTML}{df65b0}
\definecolor{pastel34}{HTML}{ce1256}

\definecolor{pastel51}{HTML}{a6611a}
\definecolor{pastel52}{HTML}{dfc27d}
\definecolor{pastel53}{HTML}{f5f5f5}
\definecolor{pastel54}{HTML}{80cdc1}
\definecolor{pastel55}{HTML}{018571}
\newcommand{\name}{\textit{EdgeWizard}}
\newcommand{\pace}{\textit{PACE}}
\newcommand{\pacecite}{~\cite{pace, pace_isocc}}

\definecolor{color1}{HTML}{FAD773} 
\definecolor{color2}{HTML}{F2BFEF} 
\definecolor{color3}{HTML}{B6F0BA} 
\definecolor{color4}{HTML}{A1DBF0} 

\tikzset{pics/wedge/.style={code={%
}},wedge/.cd,alpha/.initial=0,beta/.initial=0,color/.initial=blue,
  mix color/.initial=gray,radius/.initial=3cm,h/.initial=1cm,
/tikz/.cd,
pics/3d pie chart/.style={code={%
}},3d pie chart/.cd,
  colors/.initial={"blue","red","orange","green","yellow"},
  radius/.initial=3cm,h/.initial=1cm,alpha0/.initial=0,
  aspect/.initial=0.6,eccentricity/.initial=0.7,cname/.initial=c
}

\newtoggle{anonymized}
\toggletrue{anonymized}  

\usetikzlibrary{shadows,patterns,shapes,arrows,decorations.pathmorphing,backgrounds,positioning,fit,plotmarks,calc,spy}
\usetikzlibrary{fadings,positioning,matrix}

\begin{document}
\bstctlcite{BSTcontrol}
\title{{\name}: Motion-Resilient EEG Wearable with On-Device Processing}
\author{} 

\vspace{-2em}
\IEEEaftertitletext{%
\hbox to \textwidth{%
\begin{tabular*}{\textwidth}{@{\extracolsep{\fill}}ccc@{}}
\shortstack[c]{{\large Teck Lun Goh$^{*}$}\\
\texttt{gtecklun@comp.nus.edu.sg}\\
School of Computing\\
National University of Singapore}
&
\shortstack[c]{{\large Rohan Juneja$^{*}$}\\
\texttt{rohan@comp.nus.edu.sg}\\
School of Computing\\
National University of Singapore}
&
\shortstack[c]{{\large Vishruti Ranjan}\\
\texttt{vishruti@comp.nus.edu.sg}\\
School of Computing\\
National University of Singapore}
\\[1em]
\multicolumn{3}{c}{%
\begin{tabular}{@{}c@{\hspace{2cm}}c@{}}
\shortstack[c]{{\large Tulika Mitra}\\
\texttt{tulika@comp.nus.edu.sg}\\
School of Computing\\
National University of Singapore}
&
\shortstack[c]{{\large Li-Shiuan Peh}\\
\texttt{peh@comp.nus.edu.sg}\\
School of Computing\\
National University of Singapore}
\end{tabular}}
\\[2em]
\multicolumn{3}{c}{\footnotesize $^{*}$ These authors contributed equally.}
\end{tabular*}%
}%
}

\maketitle

\begin{abstract}
Existing EEG systems struggle with degraded signal fidelity under motion and limited responsiveness due to latency and power constraints of tethered processing - limiting their use in real-world Brain-Computer Interface (BCI) applications. We present {\name}, a fully integrated, motion-resilient EEG wearable that combines spring-loaded dry electrodes, a six-layer flexible PCB with a dedicated noise electrode layer, and a low-power SoC with a coarse-grained reconfigurable accelerator (CGRA). 

To enable efficient edge inference, we developed LPEEGNet, a 6-bit quantized, hardware-aware EEGNet variant that achieves 794x speedup and 25x memory reduction over the float32 baseline at 59.2GOPS/W. In a 15-participant IRB-approved trials, {\name} achieved 95\% Steady State Visual Evoked Potential (SSVEP) classification accuracy while users were walking at 3km/h, with end-to-end latency at 34.1ms and 71mW total power consumption (23 hours on a 500 mAh battery). 

To demonstrate the system's architectural robustness, we further validated the CGRA running LPEEGNet using the Siena Scalp EEG dataset, achieving 94.96\% balanced accuracy, nearly matching the performance of the full-precision EEGNet (95.21\%) across high-density, wet-electrode clinical data. 
\end{abstract}


\section{Introduction}
Electroencephalography (EEG) has expanded beyond traditional clinical roles to enable wearable, ambulatory applications. One prominent use case is epilepsy, which affects over 50 million people worldwide~\cite{who_epilepsy}, where continuous ambulatory monitoring could enable timely caregiver alerts and automatic event annotation during daily activities~\cite{epilepsy_intervention}. SSVEP-based brain-computer interfaces (BCIs) offer another compelling paradigm: neural responses synchronized with flickering stimuli enable robust classification using a small number of occipital electrodes, making them well-suited for wearable, low-power deployment\cite{WalkingWizard}.  Both applications share strict requirements: high EEG signal fidelity under real-world motion, low end-to-end latency, and energy efficiency for continuous operation.

Traditional signal processing techniques struggle to capture the complexity of neural signals, particularly in dynamic environments where motion artifacts degrade EEG signal quality. While computationally efficient \cite{review_1}, conventional methods lack the analytical depth required for advanced BCI tasks. Deep learning models such as EEGNet \cite{eegnet_accelerator} improve BCI accuracy by learning nonlinear neural patterns directly from raw EEG data \cite{traditional_vs_deep_learning}, but are typically evaluated under stationary conditions, revealing a critical gap for wearable, ambulatory deployment. Addressing this requires redesigned sensor interfaces, PCB architectures, and auxiliary reference channels for artifact suppression. 

On the computational side, wireless transmission of EEG data can consume up to 70\% of total system power~\cite{wearable_eeg_lit}, and delegating inference to an external workstation introduces latency exceeding 1s~\cite{WalkingWizard}. On-device inference is therefore essential for both responsiveness and sustained battery operation. 

Deploying deep learning on wearable EEG hardware is constrained by available plaforms. Commercial SoCs - such as Apple's S10~\cite{apple_neural_engine}, and Samsung's Exynos W Series~\cite{samsung_exynosW}, target general consumer applications and lack wearable EEG optimizations.
Dedicated EEG ASICs~\cite{eegnet_processor1, eegnet_processor2, eegnet_processor3} achieve microwatt-level power consumption but are tailored to narrow computational kernels, limiting flexibility for diverse BCI tasks. Additionally, they incur high development costs and require long design cycles, further restricting scalability. There is a clear need for a programmable platform tailored for real-time wearable EEG systems.

Field-Programmable Gate Array (FPGAs) can reduce EEGNet inference from 121ms to 12.9ms, but at 1.47W~\cite{cpu_gpu_fpga} which is orders of magnitude above wearable power budgets~\cite{cyton}.
Modern Coarse-Grained Reconfigurable Architectures (CGRAs)~\cite{agile_cgra} offer greater energy efficiency than FPGAs with better programmability. 
{\pace} CGRA{\pacecite} exemplifies this: operating at 4.4mW and 360~GOPS/W, unifying signal processing and neural inference in a single programmable fabric with minimal reconfiguration overhead.
\textbf{While ULP-CGRA~\cite{ulp_srp} has demonstrated viability for biomedical applications, CGRAs have yet to be validated in full end-to-end systems running deep learning models at the edge. }
To the best of our knowledge, no prior work has demonstrated CGRAs enabling on-device edge AI in a fully self-contained wearable system.

\begin{table*}[t]
    \centering
    \small
    \captionsetup{justification=centering}
    \caption{Comparison of EdgeWizard to other state-of-the-art.}
    \resizebox{\textwidth}{!} {
    \begin{tabular}{cccccccc}
        \toprule
         Related Work & 
         \begin{tabular}{c} Resistant \\ To Motion \end{tabular} & 
         \begin{tabular}{c} Fully-Dry \\ Electrodes \end{tabular} & 
         \begin{tabular}{c} Number of \\ Channels \end{tabular} & 
         \begin{tabular}{c} Power \\ (mW) \end{tabular}  &
         \begin{tabular}{c} Latency \\ (ms) \end{tabular}  &
         \begin{tabular}{c} System \\ On Chip \end{tabular} &
         \begin{tabular}{c} Application \\  \end{tabular} \\ 
         \hline

        \multicolumn{7}{c}{Wet electrode}  \\ \hline
        Chai \cite{embedded} & No & No & 2 & $>$3300 & 3000 & MCF5213 Coldfire & SSVEP \\
        McCrimmon \cite{performance_assessment_of_a} & No & No & 7  & 1000 & $>$1000 & Atmel SAM3X8E & MI \\
        Salvaro \cite{towards_a_novel_hmi} & No & No & 3  & 27.5 & 10000 & ARM-STM32F407 & SSVEP \\
        Kartsch \cite{ulp_system} & No & No & 8  & 109 & No Info & ARM-STM32F407 & Drowsiness \\
        Kartsch \cite{eeg_drowsiness} & No & No & 2  & 91.7 & No Info & ARM-STM32F407 & Drowsiness \\
        Xie \cite{review_2} & No & No & 8  & 1900-3100 & $>$1000 & Xilinx Zynq-7000 & Eyes Open / Close \\
        \hline
        
        \multicolumn{7}{c}{Dry electrode but not resistant to motion}  \\ \hline
        Salvaro \cite{embedded_cortexM4} & No & Yes & 3 &  22.4  & 2000 & ARM-STM32F407 & SSVEP \\
        Lin \cite{tablet} & No & Yes & 2  & 5000\footnotemark[1] &  $>$1000 & Qualcomm Snapdragon 800 & SSVEP \\
        Li \cite{combined_eeg_gyroscope} & No & Yes & 2  & 321.6 & $>$1000 & Qualcomm Snapdragon 400 & Alpha Power \\
        Venuto \cite{raspberrypi_eeg} & No & Yes & 32 & $>$1200 & 19.7\footnotemark[2] & Broadcom BCM2836  & P300 \\
        Chi \cite{dry_non_contact} & No & Yes & 8  & 1000 & $>$1000 & TI OMAP 2420& SSVEP \\
        Kartsch \cite{biowolf} & No & Yes & 3  & 6.31 & 1260 & Mr Wolf \cite{mrwolf} & SSVEP \\
        Woehrle \cite{fpga_wohrle} & No & Yes & 128  & 4490 & 56 & Xilinx Zynq ZC7030 & P300 \\
        \hline
        
        \multicolumn{7}{c}{Resistant to motion}  \\ \hline
        Lee \cite{mobile_bci_2} & Yes & No & 36  & $>$100,000\footnotemark[1] & $>$1000 & Intel Core i7\footnotemark[3] & SSVEP \\
        
        Mullen \cite{mobileSSVEP_9} & Yes & Yes & 64  & $>$ 100,000\footnotemark[1] & 438 & AMD Barcelona & SSVEP \\
        Goh \cite{WalkingWizard} & Yes & Yes & 4+1\footnotemark[4]  & 190\footnotemark[5] & $>$1000 & Intel Core i7\footnotemark[3] & SSVEP \\
        \hline      
         
         \begin{tabular}{c}\name \end{tabular} & Yes & Yes & 4+1\footnotemark[4]   &  71.0 & 34.0 & {\pace} SoC\cite{pace}  & SSVEP \\
         \bottomrule
    \end{tabular}}\\

    \raggedright
    \small $^1$ Total system power estimated - includes external workstation (Intel Core i7 or equivalent) running inference off-chip, typically exceeds 100W. \\
    \small $^2$ Classification only, no end-to-end system latency value provided. \\ 
    \small $^3$ Not mentioned, standard PC SoC inferred.\\
    \small $^4$ Additional electrode channel used as noise electrode to enable EEG recording using dry electrode while moving. \\
    \small $^5$ On-head power only for WalkingWizard~\cite{WalkingWizard}: ADC (54.7mW) + PIC32 MCU (82.5mW) + Bluetooth Module (52.8mW) = 190mW. Full-system power including the external workstation (Intel Core i9 + RTX 4090) exceeds 100W during inference. \\
    \label{tab:lit_review}
\end{table*}

\begin{figure}
    \centering
    \includegraphics[trim={10mm 30mm 25mm 20mm}, clip, width=\columnwidth]{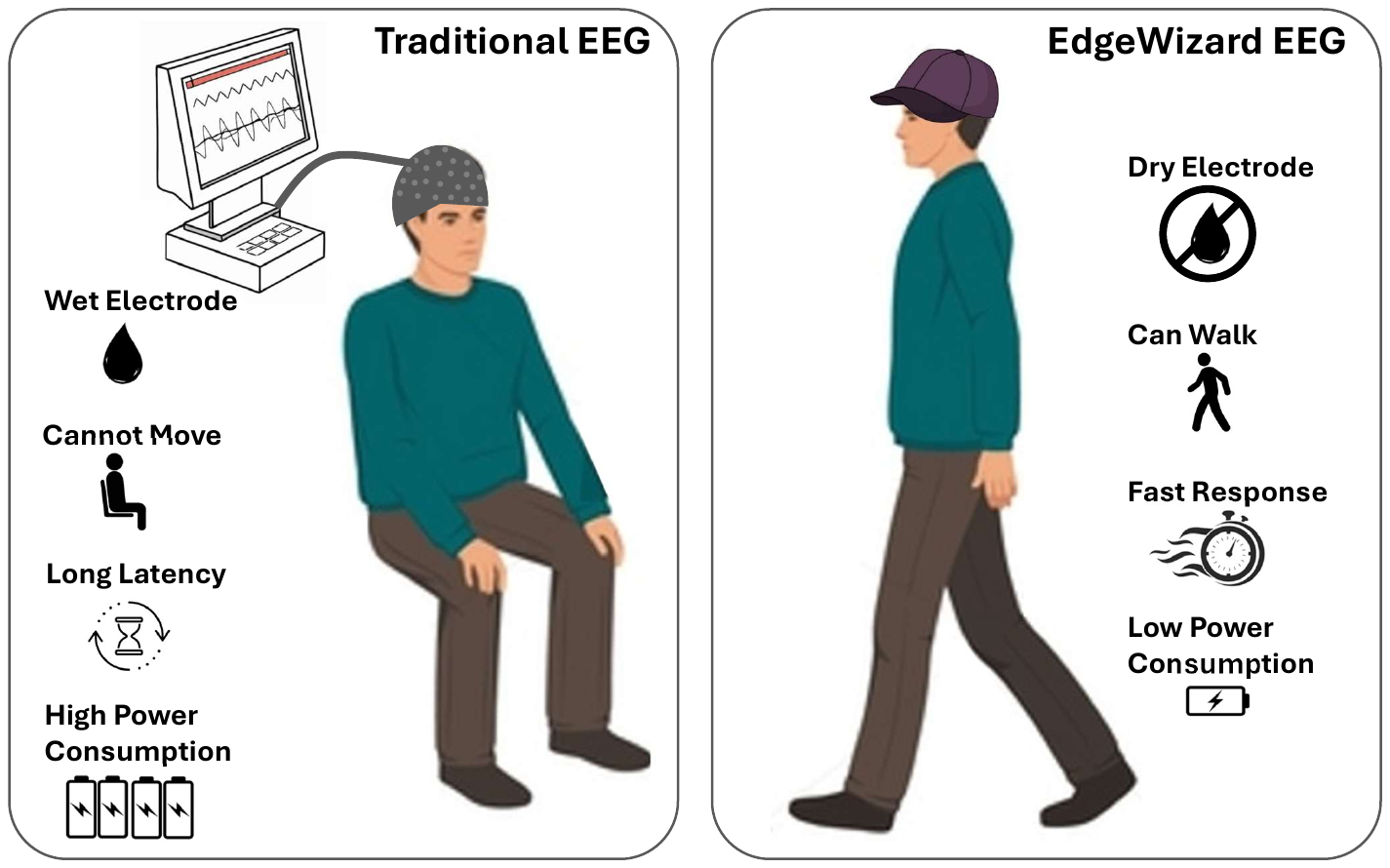}
    \caption{Traditional EEG vs. EdgeWizard. EdgeWizard uses dry electrodes, supports motion, and provides low power, real time processing for wearable BCI.}
    \label{fig:edgewizard_intro}
\end{figure}

To address this gap, we present {\name} (Fig.~\ref{fig:edgewizard_intro}), a low-power, motion-resilient EEG wearable for real-time BCI applications. {\name} integrates custom spring-loaded dry electrodes on a six-layer flexible PCB, incorporating a dedicated noise electrode layer for motion artifact suppression. {\name} is powered by {\pace} SoC running LPEEGNet, a hardware-aware quantized variant of EEGNet, entirely on-device. Validated through a 15-participant IRB-approved SSVEP user trial under standing and walking conditions, {\name} achieves 71mW total power - supporting 23 hours of battery life (500mAh), and 34ms end-to-end latency.

This paper presents the following key contributions:
\begin{itemize}
    \item \textbf{Custom SoC and hardware-aware model co-design:} The {\pace} CGRA SoC paired with LPEEGNet---an optimized EEGNet variant using 6-bit quantization, ReLU activation, and batch-norm fusion---achieves 59.2GOPS/W at 4.54mW, a 794x inference speedup, and 25x memory reduction over the float32 baseline.
    \item \textbf{Motion-resilient sensor and PCB design:} Custom spring-loaded dry electrodes feeding our CGRA SoC on a flexible PCB with a dedicated noise electrode layer, designed for robust operation. 
    \item \textbf{System-level efficiency:} On-board CGRA inference achieves 3.5x lower end-to-end latency than an STM32 CPU and 330x and 700x lower power than CPU and GPU baselines, respectively, at 34ms and 71mW total system power.
    \item \textbf{End-to-end validation:} A 15-participant IRB-approved SSVEP user study demonstrates 92\% accuracy under stationary conditions and 95\% accuracy while walking at 3km/h.
\end{itemize}

Next, Section~\ref{sec:literature_review} reviews related work and positions {\name} relative to the state of the art. Section~\ref{sec:proposed_system} describes the hardware-software co-design. Section~\ref{sec:results} presents evaluation results. Section~\ref{sec5:discussion} discusses some critical insights/ tradeoffs and Section~\ref{sec6:conclusion} concludes with limitations and future directions.
\section{Background and Related Works}
\label{sec:literature_review}
\subsection{Clinical and BCI Use Cases for Wearable EEG}
Wearable EEG holds significant potential in BCI and clinical applications. 
SSVEP - a neural response synchronized with flickering visual stimuli, enables robust classification using a small number of occipital electrodes, making it well suited for low-power wearable BCIs~\cite{ssvep_application}. Classification requires advanced methods such as canonical correlation analysis (CCA)~\cite{trca} or deep learning classifiers. Separately, continuous ambulatory EEG monitoring for epilepsy can enable real-time seizure event detection outside clinical settings - a use case explored algorithmically in this work (see Section~\ref{subsec:on_chip_performance}).

\subsection{Wearable EEG System and their Challenges}

True wearability requires dry electrodes, low power, minimal latency, and resilience despite motion~\cite{wearable_eeg_lit, category_84}. Most EEG research uses wet electrodes in controlled, stationary subjects settings~\cite{embedded, performance_assessment_of_a}, which ensures signal fidelity but prevents real-world deployment. Systems such as BioWolf~\cite{biowolf}, adopt dry electrodes for wearability but remain highly susceptible to motion artifacts.  Consumer EEG headphones~\cite{neurable, emotiv_mn8} offer a compact form factor but limited spatial resolution: for example, alpha rhythms were barely detectable on the Emotiv MN8 even in  stationary conditions~\cite{WalkingWizard}.

\cite{WalkingWizard} addressed motion artifacts through hardware-software co-design with a dedicated noise electrodes, but relies on wireless data transmission to an external workstation for inference - limiting wearability due to high latency ($>$1000ms) and power overhead (190mW on-head plus $>$100W at the workstation). 

{\name} inherits the electrode architecture of \cite{WalkingWizard} and extends it with a purpose-built SoC, hardware-aware LPEEGNet, and a fully self-contained inference pipeline - eliminating the external workstation and achieving a 59$\times$ reduction in end-to-end latency. 
Table~\ref{tab:vs_walkingwizard} lists the key differentiators and quantifies the resulting improvements.

\begin{figure*}[h]
    \centering
    \resizebox{0.95\textwidth}{!} {
    \includegraphics[width=\textwidth]{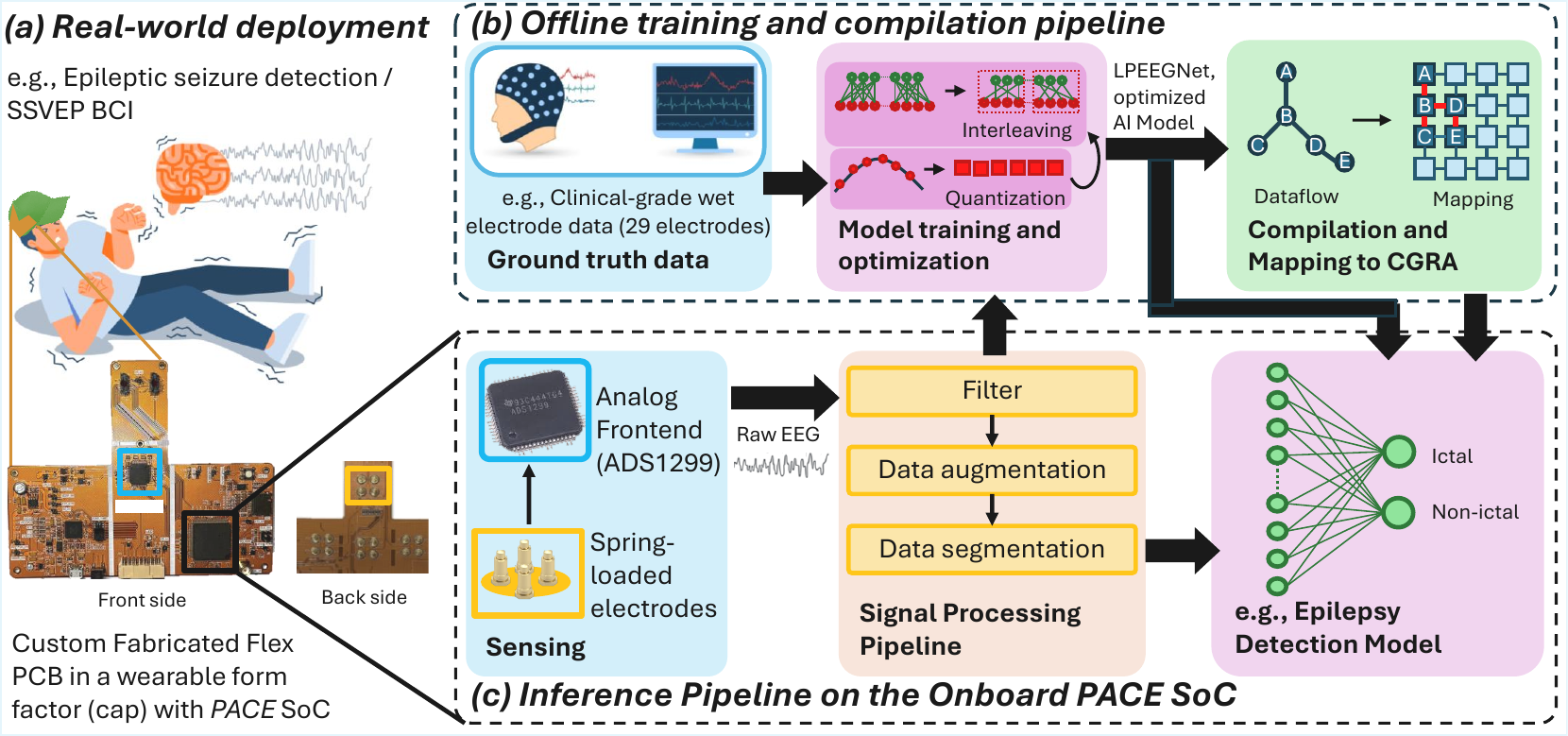}
    }
    \caption{Overview of the EdgeWizard system for real-time, motion-resilient EEG processing. (a) Real-world deployment scenario illustrating the wearable EEG system used for seizure detection. (b) Offline pipeline for training and compiling LPEEGNet, including ground-truth acquisition from wet electrodes, model optimization, and spatial mapping onto the CGRA. (c) On-device inference pipeline on the {\pace} SoC, where EEG is acquired via spring-loaded electrodes, processed in real time, and classified into cortical states.}
    \label{fig:system_architecture}
\end{figure*}

\begin{table}[h]
\centering
\caption{{\name} vs.\ WalkingWizard \cite{WalkingWizard}: Key Differentiators}
\resizebox{\columnwidth}{!}{
\begin{tabular}{lcc}
\toprule
\textbf{Feature} & \textbf{WalkingWizard \cite{WalkingWizard}} & \textbf{{\name}} \\
\midrule
Spring-loaded electrodes   & \cmark & \cmark \\
Noise electrode  & \cmark & \cmark \\
On-device inference        & $\times$ & \cmark \\
Custom SoC (PACE CGRA)     & $\times$ & \cmark \\
Hardware-aware model (LPEEGNet) & $\times$ & \cmark \\
PCB & Only ADC and electrodes & Fully integrated \\
End-to-end latency         & $>$1000~ms & \textbf{34~ms} \\
On-head power              & 190~mW     & \textbf{71~mW} \\
External workstation required & Yes     & \textbf{No} \\
Battery life (500~mAh)     & $<$9~hr    & \textbf{23~hr} \\
\bottomrule
\end{tabular}}
\label{tab:vs_walkingwizard}
\end{table}

\subsection{EEG Signal Processing and Hardware Platforms}
EEG signals are inherently low-amplitude and highly susceptible to noise, with motion artifacts further degrading quality in dynamic environments. Traditional signal processing (e.g. CCA, FFT-based thresholding) is computationally efficient \cite{review_1} but loses accuracy during movement~\cite{jose_museum_51}. Deep learning, particularly EEGNet \cite{eegnet} - a compact CNN using depthwise and separable convolutions significantly improves classification by learning nonlinear patterns from raw data~\cite{traditional_vs_deep_learning}. Prior works apply model compression and quantization techniques~\cite{q-eegnet, efficient_model_compressed_eegnet} but address inference only, without an integrated EEG acquisition and processing system.

As shown in Table~\ref{tab:lit_review}, most studies \cite{embedded, performance_assessment_of_a, towards_a_novel_hmi, ulp_system, eeg_drowsiness, review_2} use wet electrodes and restrict user movement to avoid motion artifacts and rely on lightweight SoCs (e.g. ARM-STM32F407) that support only basic FFT-based algorithms. Dry-electrode systems~\cite{embedded_cortexM4, tablet, combined_eeg_gyroscope, raspberrypi_eeg, dry_non_contact, biowolf} remove the need for conductive gel but are highly susceptible to motion artifacts \cite{dry_electrode}. EEG systems that are able to handle motion artifact~\cite{mobile_bci_2, mobileSSVEP_9, WalkingWizard} demonstrate ambulatory EEG  feasibility but transmit data wirelessly to external workstations, incurring high latency and system power overhead. 

Achieving this on a wearable power budget demands both energy efficiency and programmability. Recent hardware solutions~\cite{eyeriss, pudianNao, react, nova} have turned to microarchitectural specialization to maximize energy efficiency, but at the cost of programmability, and high non-recurring engineering costs. Given the rapidly evolving nature of AI in the wearable domain, such premature specialization risks obsolescence as models and tasks continue to advance. {\name} is the first fully self-contained, dry-electrode wearable to perform on-device deep learning inference under motion, eliminating the external workstation while achieving lower power and latency than any prior motion-capable system.  
\section{Proposed System} 
\label{sec:proposed_system}
\subsection{System Design Overview}
\textbf{Real-world application.} 
{\name} is designed to support a broad range of EEG-based applications. In this work, we illustrate its application on  two representative EEG use cases: (1) an SSVEP-based BCI, evaluated end-to-end with 15 participants under standing and walking conditions, and (2) epileptic seizure detection, evaluated at the algorithmic level using the publicly available Siena Scalp EEG dataset. The system-level design is illustrated in Fig.~\ref{fig:system_architecture}.

\textbf{Overview of hardware-software stack.}
EdgeWizard integrates sensing and computation within a compact wearable platform designed for real-time 
EEG processing in BCI applications. 
The entire SoC design along with the software stack and PCB are developed in-house.
As shown in Fig.~\ref{fig:system_architecture}, the system-level integration includes dry electrodes, the analog frontend with ADC, and the {\pace} SoC assembled on a flexible PCB. The software stack leverages the {\pace} compiler to deploy LPEEGNet, a quantization-optimized variant of EEGNet, through calibration, quantization, and static mapping of model layers onto the CGRA. This hardware-software co-design enables EdgeWizard to support motion-resilient, fast EEG inference suitable for continuous deployment at low power.

\begin{figure*}[h]
    \centering
    \resizebox{\textwidth}{!} {
    \includegraphics[width=\linewidth]{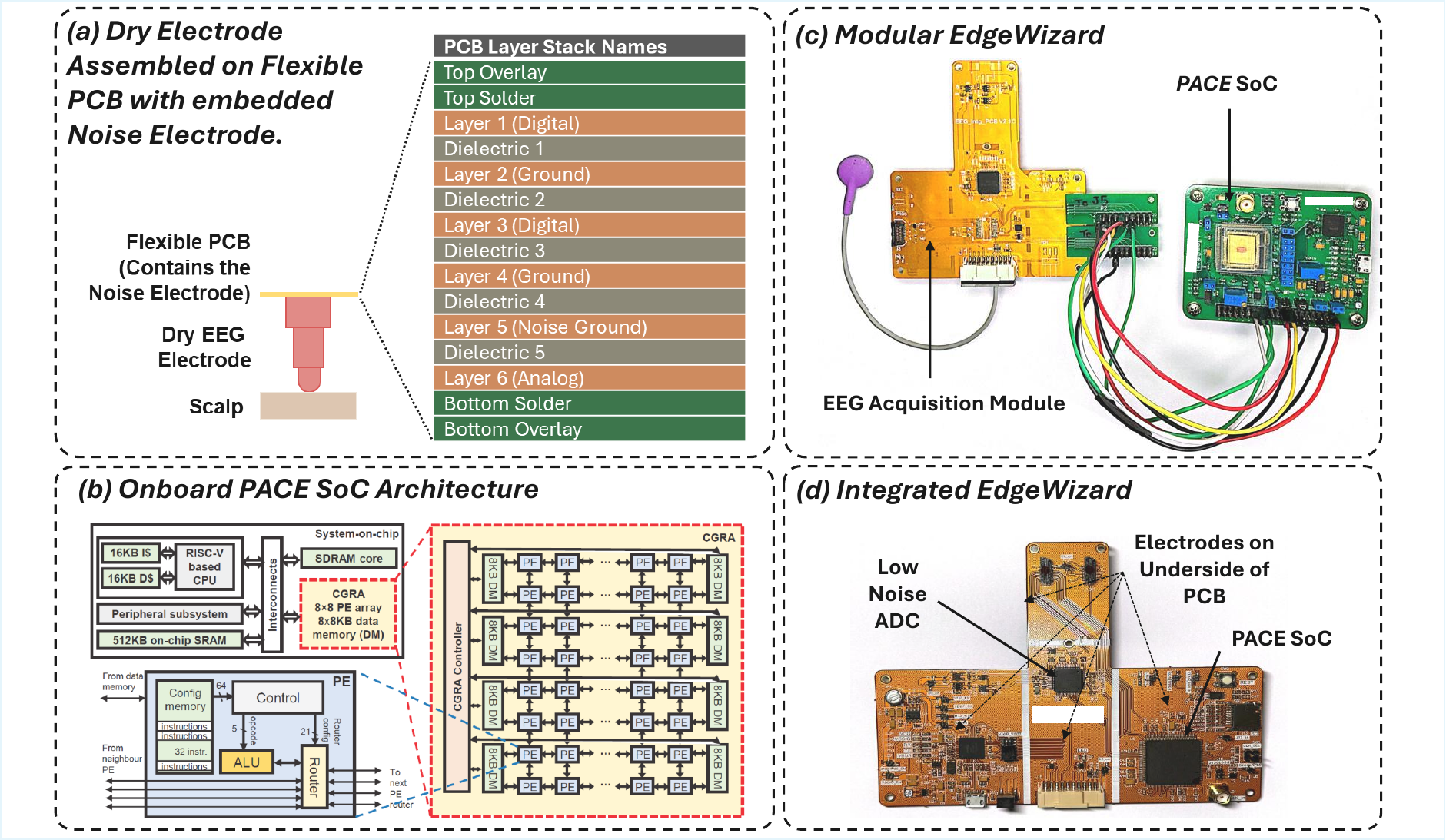}
    }
    \caption{EdgeWizard's hardware components were custom-designed, from the electrodes to the {\pace} CGRA accelerator chip, and the fully integrated PCB.}
    \label{fig:hw}
\end{figure*}

\subsection{Motion-Resilient Sensor and PCB Design}
\textbf{Sensor design and analog frontend.}
To ensure high-fidelity signal capture over hair-covered scalp regions, electrodes are designed in a combed structure - using four spring-loaded pins for each electrode. We used the MIL-MAX 0871-0-57-20-82-14-11-0 \cite{mill_max} spring-loaded pin due to the following attributes: (a) gold-plating ensures bio-compatibility with the scalp, (b) 2mm stroke length to accommodate minor electrode-scalp displacement to minimize impedance fluctuation, (c) 10mm pin length to penetrate dense or long hair, and (d) relatively large diameter to maximize contact area and minimize impedance. The EEG analog frontend utlizes the Texas Instruments ADS1299, an 8-channel, low-noise, high-resolution ADC with integrated programmable gain amplifier \cite{ads1299}. The ADC interfaces with a custom {\pace} SoC via an SPI bus, thereby facilitating on-device signal processing (architecture details in Section~\ref{subsec:pace_soc}). To streamline prototyping and validation, the hardware development followed a two-phase approach:
\begin{itemize}
\item \textbf{Phase 1: Modular Development} – The {\pace} SoC and the EEG acquisition module were fabricated as separate PCBs (Fig.~\ref{fig:hw}C). These were interconnected to allow isolated testing of individual subsystems and integration testing of full system functionality. This configuration is referred to as Modular EdgeWizard.
\item \textbf{Phase 2: Fully Integrated System} – The final EdgeWizard platform (Fig.~\ref{fig:hw}D) consolidates both modules into a single flexible PCB, reducing wiring complexity and improving mechanical robustness. This configuration is referred to as Integrated EdgeWizard.
\end{itemize}

\textbf{PCB design and fabrication.}
Integrated EdgeWizard is built on a six-layer flexible PCB. 
To suppress motion artifacts, we adapt techniques from \cite{WalkingWizard} by embedding a dedicated noise electrode layer (Fig.~\ref{fig:hw}A, Layer 5) within the PCB. The PCB is designed with the electrodes assembled at POz, Oz, O1 and O2 electrode position, and connected to the first 4 channels of the ADC. 
The fifth ADC channel is allocated to the noise electrode layer - isolated from cortical sources and captures motion-induced artifacts \cite{daniel_isolate_15, daniel_dual_electrode_19, daniel_dual_electrode_24}, which aids in subsequent artifact rejection. The remaining three channels of the ADC are routed to a 10-pin connector (Fig.~\ref{fig:hw}D), reserved for future expansion and debugging. 

Given the low amplitude of EEG signals and their susceptibility to interference from high-speed digital traces, careful consideration is given to signal routing and layer allocation. All analog EEG signals are routed on the bottom-most layer (Layer 6) to isolate them from high-speed digital lines. Digital signals are confined to Layers 1 and 3, while Layers 2 and 4 serve as digital ground planes to suppress signal loops and electromagnetic emissions. This separation of analog and digital domains ensures that a clean EEG signal can be recorded even in the presence of high-speed digital activity.

\subsection{Data Acquisition and Model Development}
\label{subsec:data_acq_model_development}
\textbf{Ground truth annotation:}
For the epilepsy use case, direct testing on epileptic patients using {\name} was not feasible. Instead, we evaluated the performance of LPEEGNet using the publicly available Siena Scalp EEG Dataset \cite{siena_scalp} as an exploratory benchmark. 
This dataset comprises EEG recordings from 14 patients, with seizure events annotated by clinicians. The raw EEG were split into 60\% training, 20\% validation and 20\% test sets. 

\begin{tcolorbox}[
  colback=yellow!8, colframe=orange!70!black,
  boxrule=1pt, arc=2pt,
  left=4pt, right=4pt, top=3pt, bottom=3pt,
  title={\small\textbf{Note}}]
{\small The Siena dataset uses \emph{wet-electrode} EEG at 512~Hz across 29 channels; {\name} records \emph{dry-electrode} EEG at 250~Hz across 5 channels. The epilepsy results in Section~\ref{subsec:on_chip_performance} characterize \textbf{LPEEGNet's algorithmic capability} on a representative public benchmark, \emph{not} {\name}'s end-to-end system performance. Clinical validation would require in-situ patient recordings with {\name} hardware.}
\end{tcolorbox}

For the SSVEP use case, user trial was conducted with 15 participants. Details of the protocol and data collection are provided in Section~\ref{sys_performance}. 

\textbf{Training data summary:} LPEEGNet for SSVEP was trained exclusively on Stage 1 participant data (80/20 train/validation split). LPEEGNet for the Siena epilepsy benchmark was trained and evaluated entirely within the Siena dataset (60/20/20 train/validation/test split). No cross-dataset training was performed.

\textbf{EEG Pre-processing:}
Following acquisition, the digitized EEG signals are processed through a structured digital signal processing pipeline that enhances signal quality and preserves features essential for inference. A bandpass filter (BPF) is applied to retain neural activity relevant to the respective tasks while removing low-frequency drift and high-frequency noise. A 50Hz notch filter is used to eliminate power line interference. 

To improve generalization and robustness to variability introduced by motion or sensor noise, the training data is augmented using several techniques: additive Gaussian noise to simulate sensor-level fluctuations and amplitude scaling to emulate signal gain variability. These transformations expand the training distribution without requiring additional data collection. 

The augmented signals are then segmented into fixed-length temporal windows using a sliding window approach, with each segment spanning 3 seconds, and downsampled to reduce computational load while retaining key frequency components. The overall processing pipeline, including filtering, augmentation and segmentation, is depicted in Fig.~\ref{fig:system_architecture}. The resulting signals are fed as input to LPEEGNet. 

\begin{figure}
    \centering
    \includegraphics[width=\columnwidth]{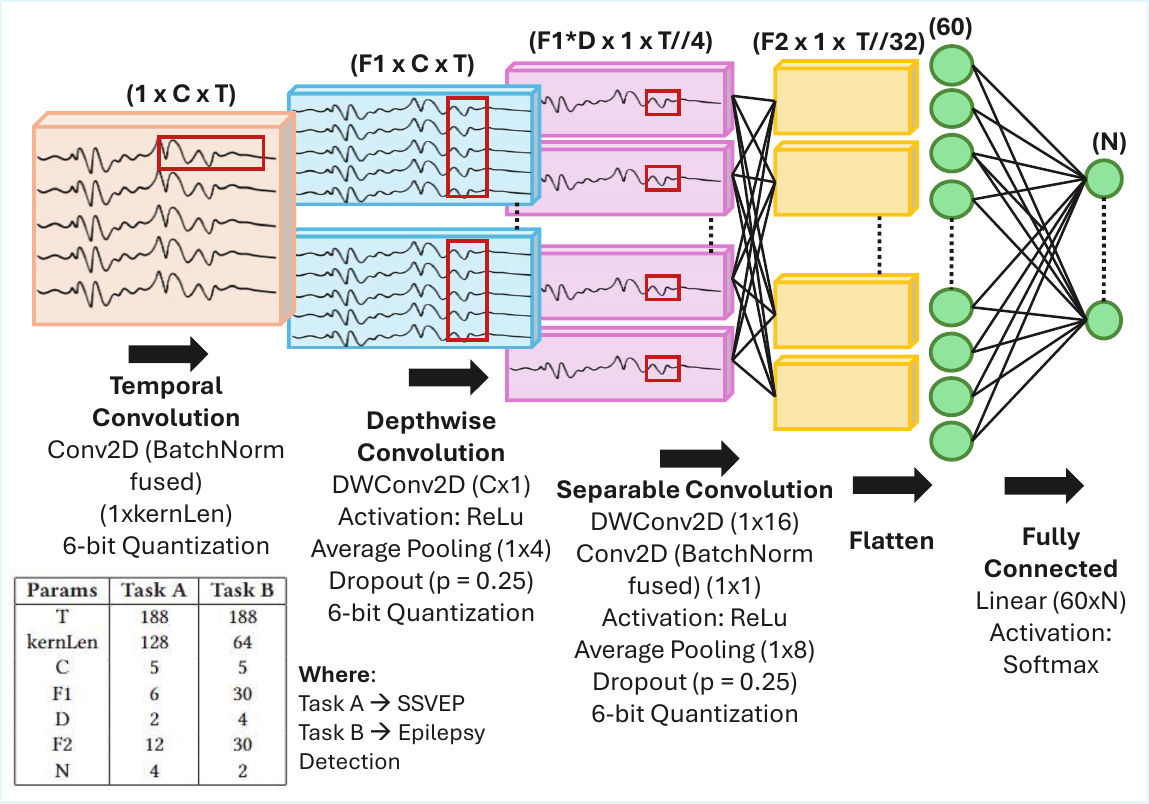}
    \caption{Overview of LPEEGNet. Raw EEG data is processed through bandpass and notch filters before inference. Parameters are adjusted for different applications, as specified in the included table.}
    \label{fig:lpeegnet}
\end{figure}

\textbf{Model training and optimization:}
\label{subsec:sw_implementation}
LPEEGNet (see Fig.~\ref{fig:lpeegnet}) was trained offline on preprocessed data collected. Training employed the Adam optimizer with an initial learning rate of $5*10^{-4}$, a batch size of 32, and categorical cross-entropy loss over 400 epochs, with the learning rate decayed by a factor of 0.5 upon validation plateaus. The model checkpoint with the highest validation accuracy was selected for subsequent quantization, optimizations and hardware mapping.

\renewcommand{\thesubsubsection}{\Alph{subsubsection}}
\subsubsection{Model Quantization:}
Post-Training Quantization (PTQ) was applied to reduce model precision from floating-point to 6-bit integer representation, thus lowering model size, memory bandwidth, and computational overhead while maintaining inference accuracy. The {\pace} CGRA processes 6-bit activations and weights within its 16-bit datapath, ensuring compatibility the new inference bitwidth \cite{nvidia_cuda_math_api}. Outputs are accumulated at 16-bit precision to preserve numerical stability before truncation to 6-bit for off-chip storage.
\begin{equation} 
Y_q = \text{clip} \left( \sum_{i=1}^{N} \left( W_i^q \cdot X_i^q \right) + B^q, a_{\min}, a_{\max} \right) 
\end{equation}

where \(Y_q\) is the quantized output, \( W_i^q \) and \( X_i^q \) are 6-bit quantized weights and activations, \(B^q\) is the quantized bias, and \(\text{clip}\) ensures values remain within \([a_{min}, a_{max}]\), preventing overflow.

This quantization strategy reduces memory latency, computational complexity, and power consumption by minimizing switching activity. Experimental results confirm that 16-bit accumulation introduces no measurable accuracy loss, ensuring LPEEGNet maintains high inference performance within the energy constraints of wearable EEG applications.

\subsubsection{Neural Network Optimizations:}
To improve computational efficiency, several optimizations were applied to EEGNet. The original model used Exponential Linear Unit (ELU) activations, which require expensive exponentiation. These were replaced with Rectified Linear Unit (ReLU) activations, which apply simple thresholding:
\begin{equation}
\text{ReLU}(x) = \max(0, x)
\end{equation}

Batch normalization was fused with convolution layers to eliminate redundant memory accesses. Given batch normalization applied to an input \(x\):
\begin{equation}
\hat{x} = \frac{x - \mu}{\sigma}, \quad y = \gamma \hat{x} + \beta
\end{equation}
where \(\mu\), \(\sigma\) are the batch mean and standard deviation, and \(\gamma\), \(\beta\) are learnable scaling and shifting parameters. These were merged into the convolution parameters:
\begin{equation}
W' = \frac{W \gamma}{\sigma}, \quad b' = \beta - \frac{\mu \gamma}{\sigma}
\end{equation}
removing the need for explicit batch normalization layers, reducing execution latency and memory overhead.

To further optimize memory usage, temporal and depthwise convolutions were interleaved, reducing intermediate storage by 85\%. Instead of storing intermediate results between layers, a single spatial column is processed at a time, immediately followed by its corresponding depthwise convolution. This localized computation minimizes memory accesses, storing intermediate results in the same memory location, significantly reducing bandwidth usage and overall memory footprint.

\subsection{{\pace} CGRA Architecture and Model Mapping}
\label{subsec:pace_soc}
\begin{table}[h]
    \centering
    \resizebox{\columnwidth}{!} {
    \begin{tabular}{|c|c|}
        \hline
        \textbf{Features} & \textbf{Value} \\\hline
        
        Technology & UMC 40nm LP CMOS \\
 Area ($mm^2$)&3.02\\
         Num of PEs& 64\\
        \multirow{2}{*}{\begin{tabular}[c]{@{}l@{}}Memory\end{tabular}}&64 KB (CGRA Data Memory)\\
 &512 KB (On-chip SRAM)\\
         Peak Operating Frequency (MHz)& 100\\
 Peak Power efficiency (GOPS/W)&360\\
         
 Peripheral Interfaces&SPI, I2C, GPIO\\\hline
    \end{tabular}
    }
    \caption{{\pace} SoC Specifications}
    \label{tab:pace_soc}
\end{table}
We designed the {\pace} SoC which pairs an open-source RISC-V Rocket core with an 8×8 CGRA accelerator optimized for energy-efficient execution, enabling high-performance inference at the edge under stringent power constraints. The RISC-V core which serves as a controller was selected for its low power footprint and broad system software support. Optimized for EEG signal processing and AI inference, it efficiently executes preprocessing tasks (bandpass and notch filtering) and deep learning inference (LPEEGNet) on-chip, reducing latency and power consumption while achieving 360GOPS/W at 4.4mW, making it ideal for wearable BCIs in dynamic conditions. As shown in Fig.~\ref{fig:hw}B, the {\pace} SoC features an optimized dataflow architecture with a hierarchical memory system to enhance performance and energy efficiency. The RISC-V processor manages system control and task scheduling, while the CGRA executes computational kernels in parallel, reducing processing bottlenecks. A multi-hop on-chip network enables low-latency communication, and each processing element (PE) contains 8 KB of local memory, minimizing off-chip accesses. Additionally, 512KB of shared SRAM supports larger dataset handling. Fabricated in a 40nm LP process, the {\pace} SoC is well-suited for energy-constrained BCI applications, with its key specifications summarized in Table~\ref{tab:pace_soc}. To optimize power consumption, {\pace} incorporates idle state instructions and dynamic clock gating. Unused PEs enter a low-power state using No Operation (NOP) instructions, significantly reducing energy consumption during inactive cycles. This fine-grained power management dynamically adjusts to workload variations, ensuring maximum efficiency for real-time, low-power EEG processing.

The {\pace} compiler first uses LLVM to lower each signal processing or neural network kernel into a single static assignment intermediate representation (IR). It then converts that IR into a directed dataflow graph that explicitly encodes both computations and their data dependencies. A simulated annealing mapper assigns each graph node to a PE and determines routing paths to minimize the initiation interval, which refers to the number of cycles between successive outputs and directly impacts performance, while also balancing on-chip memory usage. Finally, the compiler generates the configuration bitstream required by the CGRA fabric, enabling deeply pipelined and parallel execution of filters and LPEEGNet layers with minimal control overhead~\cite{agile_cgra},~\cite{nexus},~\cite{snafu},~\cite{flex}. 

\subsection{Online Inference and Firmware Integration}
\lstdefinestyle{api}{
  language=C,
  basicstyle=\ttfamily\footnotesize,
  keywordstyle=\color{blue}\bfseries,
  commentstyle=\color{gray},
  stringstyle=\color{red},
  numbers=left,
  numberstyle=\tiny,
  numbersep=5pt,
  breaklines=true,
  frame=single,
  captionpos=b,
  morekeywords={HAL_PACE_Init,HAL_PACE_Read,HAL_PACE_Write,HAL_PACE_Flush,HAL_PACE_Ctrl_Read,HAL_PACE_Ctrl_Write,HAL_PACE_Config_Read,HAL_PACE_Config_Write}
}

\begin{lstlisting}[style=api, caption={Key functions for configuring and interacting with the CGRA in the {\pace} SoC.}, label={lst:firmware}]
typedef struct {
  volatile unsigned int* Core;       // CGRA memory controller registers      
  volatile unsigned int* Control;  // CGRA control registers
} HAL_PACE_TypeDef;

// Initializes the CGRA interface.
void HAL_PACE_Init(HAL_PACE_TypeDef* hinst);

// Accesses CGRA control registers (reset, enable, start, stop).
unsigned int HAL_PACE_Config_Read (HAL_PACE_TypeDef* hinst);
void HAL_PACE_Config_Write (HAL_PACE_TypeDef* hinst, unsigned int const data);
// Flushes CGRA memory with a constant.
void HAL_PACE_Flush(HAL_PACE_TypeDef* hinst, unsigned int address, const unsigned int data, unsigned int length);

// Accesses CGRA memory-mapped registers or data buffers.
void HAL_PACE_Read(HAL_PACE_TypeDef* hinst, unsigned int address, unsigned int data[], unsigned int length);
void HAL_PACE_Write(HAL_PACE_TypeDef* hinst, unsigned int address, const unsigned int data[], unsigned int length);
\end{lstlisting}

\textbf{Firmware:} The firmware, implemented in C on the RISC-V Rocket core, interfaces with the ADS1299 ADC via a hardware abstraction layer (HAL) providing CGRA initialization, configuration read/write, memory flush, and data transfer routines. The {\pace} SoC system clock runs at 12.5MHz with a clock divider setting the SPI clock (SCLK) to 781.25kHz. The ADC is configured to sample at 250SPS, sufficient to capture EEG signals up to 100Hz per the Nyquist criterion~\cite{nyquist_for_eeg}, covering the relevant clinical EEG frequency range~\cite{eeg_common_freq}.

\textbf{Real-time signal processing and inference execution on {\pace} SoC:}
After digital preprocessing, the segmented EEG data is passed to the LPEEGNet model for inference on the {\pace} SoC. Execution is performed on the CGRA, where HAL API calls (Listing~\ref{lst:firmware}) are used to transfer compiled configuration files from {\pace} compiler to the CGRA's memory system. The RISC-V core subsequently manages CGRA control via configuration registers, enabling efficient layer-wise execution through coordinated data movement and control signaling. This mapping enables dataflow-style execution within each layer, allowing intermediate results to propagate directly between compute units through local interconnects, thereby avoiding global memory transfers. Layers such as temporal convolutions, depthwise operations, and activation functions are executed sequentially, with parallelism exploited spatially across input channels and filters. The RISC-V core manages control flow and orchestrates layer-wise execution by configuring the CGRA, initiating computation, and handling data transfers between layers. Once inference completes, the classification output is collected by the RISC-V core and used to update the application layer, which may include triggering real-time feedback to the user. This architecture enables low-latency and energy-efficient inference suitable for continuous operation in wearable BCI systems.  
\section{Evaluation}
\label{sec:results}
\subsection{EEG Analog Front-End Characterization}
\begin{figure}[h]
    \centering
    \includegraphics[width=0.75\linewidth]{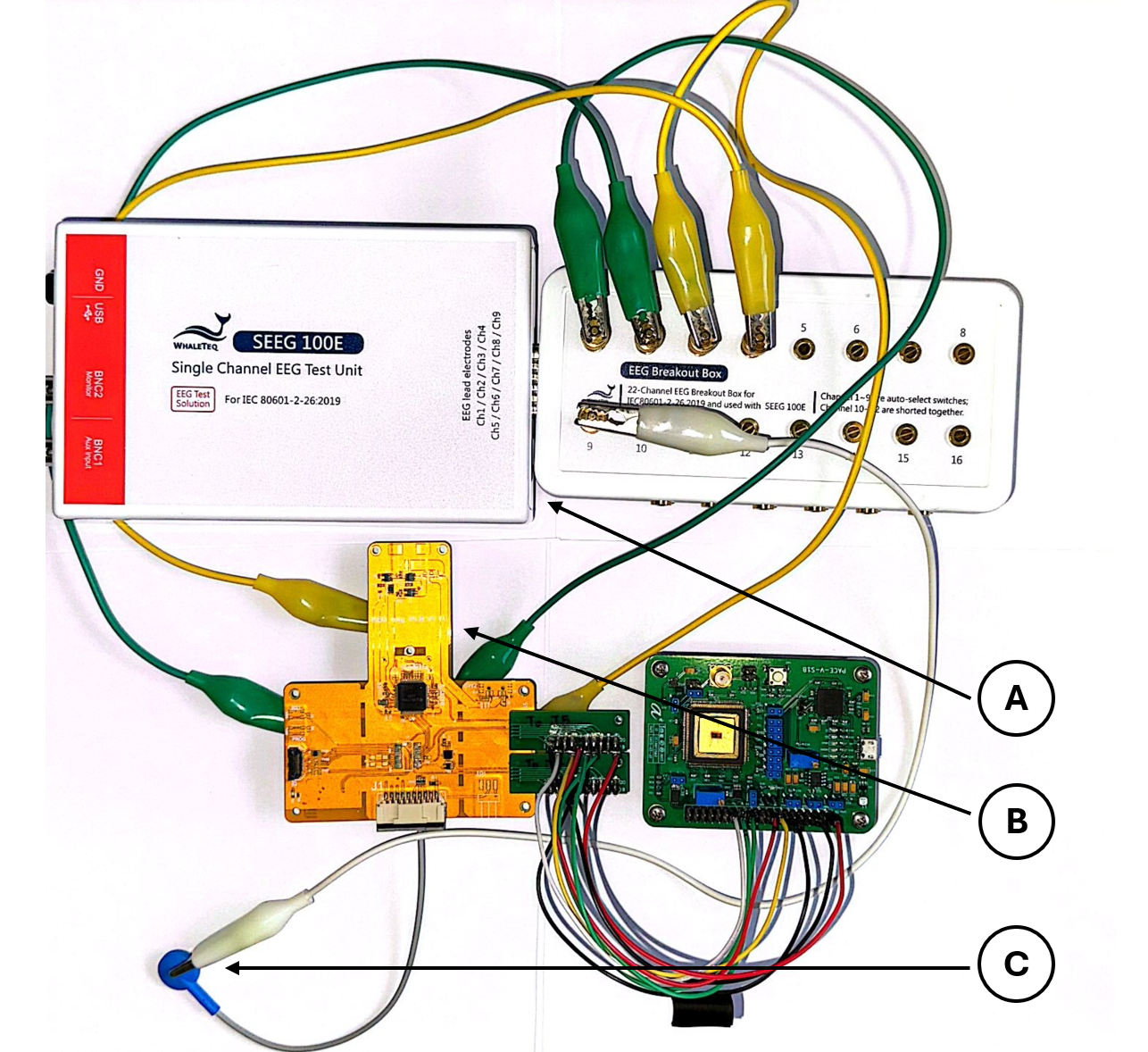}
    \caption{EEG Analog Front-End Characterization Setup. (A) SEEG100E was used to generate the test stimulus for the characterization. (B) Electrodes on underside of Modular EdgeWizard connected to channels 1 to 4 of SEEG100E. (C) Reference electrode to channel 9 (Ground) of SEEG100E.}
    \label{fig:seeg100e}
\end{figure}

\begin{table}[ht]
  \centering
  \resizebox{\columnwidth}{!}{
  \begin{tabular}{|c|c|c|}
    \hline
    \textbf{Test} & \textbf{Specified limits} & \textbf{Measured}\\
    \hline
    Amplitude Accuracy
      & $\pm 20\%$ of reference
      & {\cellcolor[HTML]{C8E6C9}}{1.01--1.02\,mV (1mV input)}\\
    Dynamic Range
      & \begin{tabular}{c}
      $\pm 10\%$ of 1mV input \\
      with $\pm 150$\,mV DC offset
      \end{tabular}
      & {\cellcolor[HTML]{C8E6C9}}{$\pm 9.5\%$ of 1mV input}\\
    Input-Referred Noise
      & $<6\,\mu\mathrm{V}$\,peak--to--peak
      & {\cellcolor[HTML]{C8E6C9}}{$0.75\,\mu\mathrm{V}$\,peak--to--peak}\\
    Frequency Response
      & 71--110\% over 0.5--50\,Hz
      & {\cellcolor[HTML]{C8E6C9}}{95--102\%}\\
    \hline
  \end{tabular}
  }
  \caption{Summary of key EEG analog front-end performance metrics measured on both Modular and Integrated EdgeWizard. Tests conform to IEC 80601-2-26:2019 and include amplitude accuracy, dynamic range, input-referred noise, and frequency response. All measured values fall within their specified limits.
  }
  \label{tab:key_electrical_metrics}
\end{table}

\textbf{Methodology:} 
To verify EEG acquisition fidelity we subjected the EdgeWizard frontend to the IEC 80601-2-26 2019 EEG Performance standard \cite{iec80601-2-26}, satisfying a key prerequisite for medical certification. Testing used the SEEG100E system \cite{seeg100e} to produce the required test waveforms. Table~\ref{tab:key_electrical_metrics} summarizes the measured results for Modular and Integrated EdgeWizard. All metrics meet or exceeed the standard's pass criteria. 

\subsubsection{Accuracy of Amplitude and Rate of Variation (§201.12.1.102):}
Test stimulus: 6~Hz triangular waveform at 1, 0.5, 0.2, and 0.1~mV. Pass criterion: recorded amplitude within $\pm$20\% of stimulus.

\subsubsection{Input Dynamic Range and Differential Offset Voltage (§201.12.1.103):}
Test stimulus: 6~Hz triangular waveform at 1~mV with DC offsets of $-$150, 0, and $+$150~mV. Pass criterion: recorded amplitude between 0.9--1.1~mV regardless of DC offset.

\subsubsection{Input Noise (§201.12.1.104):}
Test stimulus: all EEG channels shorted to reference. Pass criterion: peak-to-peak noise below 6~$\mu$V.

\subsubsection{Frequency Response (§201.12.1.105):}
Test stimulus: 666~$\mu$V sinusoid at 0.5, 5, and 50~Hz. Pass criterion: recorded amplitude between 71--110\% (473--733~$\mu$V) of stimulus.

EdgeWizard passes all four tests (Table~\ref{tab:key_electrical_metrics}), confirming linear amplitude fidelity, $\pm$150~mV DC offset tolerance without signal distortion, a 0.75~$\mu$V noise floor (8$\times$ below the 6~$\mu$V limit), and flat frequency response (95--102\%) across the clinically relevant 0.5--50~Hz band.

\subsubsection{Impedance Characteristics Compared to Wet Electrodes:}
Electrode-scalp impedance is a critical indicator of contact quality and expected signal quality prior to EEG acquisition. We compare the impedance of EdgeWizard dry electrode against (i) wet electrode without skin preparation, and (ii) OpenBCI combed dry snap electrode. 
The wet electrode was applied using Ten20 Conductive Paste without prior skin abrasion. 

\begin{figure}[h]
    \centering
    \resizebox{\columnwidth}{!} {
    \includegraphics[width=\linewidth]{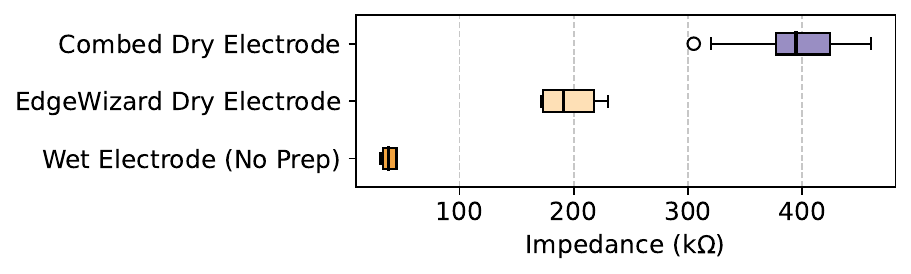}
    }
    \caption{Comparison of electrode-scalp impedance for wet electrode, EdgeWizard dry electrode, and a combed dry electrode (OpenBCI).}
    \label{fig:impedance}
\end{figure}

As shown in Fig.~\ref{fig:impedance}, while {\name} electrode-scalp impedance is higher than a wet electrode but significantly lower than a standard combed dry electrode (OpenBCI), falling at the lower end of the range reported for dry contacts \cite{electrode_impedance}. This reduced impedance contributes to improved signal quality and motion resilience, as evidenced by the dry-wet correlation analysis in Section~\ref{subsec:motion_resilience}.

\subsection{On-Chip Compute Performance}
\label{subsec:on_chip_performance}
\textbf{Methodology:} To evaluate the standalone performance of the {\pace} SoC, we deployed LPEEGNet on two BCI tasks: four-class SSVEP classification and epilepsy detection (binary classification). For the SSVEP task, in-house data were collected during controlled experiments involving participants performing a four-class visual stimulation task while standing and walking at speeds up to 5 km/h. The EEG pre-processing steps described in Section~\ref{subsec:data_acq_model_development} were applied, with BPF setting from 5 to 30Hz, step-size of 0.2s for the sliding window augmentation, and downsample of factor 4. For epilepsy detection, we used the publicly available Siena Scalp EEG Database \cite{siena_scalp}, which contains 29-channel EEG recordings sampled at 512 Hz from 14 patients following withdrawal of antiepileptic medication. The EEG pre-processing steps described in Section~\ref{subsec:data_acq_model_development} were again applied, with BPF setting from 0.5 to 80Hz, and downsample of factor 8. To address the strong class imbalance, seizure segments were oversampled using sliding windows with step-size of 1.5s, and augmented with Gaussian noise at four magnitudes to increase seizure samples fivefold; non-seizure segments were undersampled using step-size of 5s. 

The baseline EEGNet model was executed in 32-bit floating point on the RISC-V Rocket core, as the CGRA supports 16-bit operations. The quantized LPEEGNet was compiled to the CGRA using the {\pace} compiler, which automates model quantization, instruction generation, and spatial mapping. Inference latency was recorded during on-chip execution, and power consumption was measured by tapping the VDD rail of the evaluation board.

\subsubsection{Accuracy on SSVEP Task and Exploratory Epilepsy Benchmark:}
\begin{figure}[h]
	\centering
	\includegraphics[width=\columnwidth]{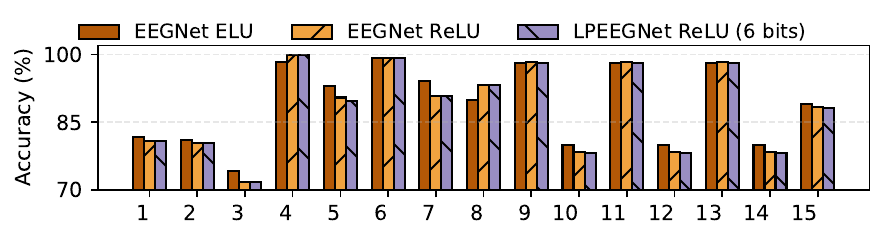}
    \caption{Offline SSVEP classification accuracy (\%) across 15 subjects using Tensorflow. EEGNet (ELU) achieved $89.0 \pm 8.4\%$, EEGNet (ReLU) $88.4 \pm 9.3\%$, and 6-bit LPEEGNet (ReLU) $88.3 \pm 9.3\%$.}
    \label{fig:ssvep_accuracy}
\end{figure}
\begin{figure}[h]
	\centering
	\includegraphics[width=\columnwidth]{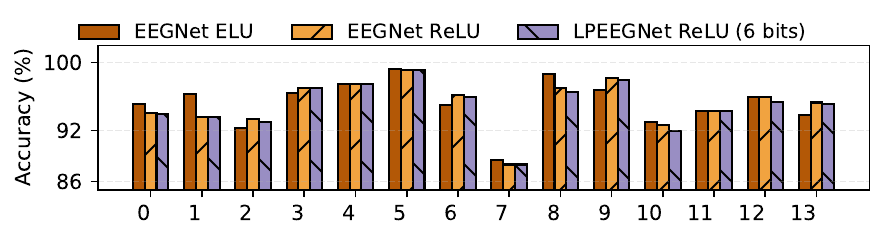}
    \caption{Balanced binary classification accuracy (\%) across 13 subjects of the Siena Dataset \cite{siena_scalp} using Tensorflow. EEGNet (ELU) achieved $95.21 \pm 2.68\%$, EEGNet (ReLU) $95.16 \pm 2.73\%$, and 6-bit LPEEGNet (ReLU) $94.96 \pm 2.74\%$. Balanced accuracy is used as the dataset is extremely unbalanced. Balanced accuracy is calculated by taking the mean of the sensitivity and specificity.}
    \label{fig:epilepsy_accuracy}
\end{figure}

As summarized in Fig.~\ref{fig:ssvep_accuracy} and Fig.~\ref{fig:epilepsy_accuracy}, LPEEGNet preserves high classification accuracy across both tasks, demonstrating minimal degradation relative to the baseline EEGNet in 32-bit floating point. For SSVEP classification, LPEEGNet achieved an average accuracy of 88.3\%, closely matching the baseline's 89.0\%. For the exploratory epilepsy detection benchmark (Siena dataset: \emph{wet electrodes, 512~Hz, 29 channels}---hardware-mismatched from {\name}'s 5 dry channels at 250~Hz), LPEEGNet achieved 91.5\% sensitivity and 98.4\% specificity (balanced accuracy: 95.0\% vs 95.2\% for the full-precision model) on matched-format inputs. This demonstrates algorithmic potential for seizure-relevant classification pending direct hardware validation. \textbf{These results demonstrate that aggressive quantization to 6-bit precision can still retain competitive accuracy, validating LPEEGNet as a viable low-power model for low-latency, on-device EEG classification.}

\subsubsection{Performance and Energy Efficiency Evaluation of Hardware-Aware Optimizations:} 
\begin{table}[h]
    \centering
    \resizebox{\columnwidth}{!} {
    \begin{tabular}{|l | c | c | c |}
        \hline
        & \textbf{Baseline} & \textbf{Opt A} & \shortstack{\textbf{Opt B}\\(LPEEGNet)}\\
        \hline
        Temporal Conv. [ms]& 12660& 24.65& 14.44\\
        Spatial Conv. [ms]& 270& 1.75& 0.90\\
        Spatial Conv. [ms]& 154& 1.38& 0.72\\
        Sep. Conv. [ms]& 119.8& 0.91& 0.54\\
        FC [ms] & 3.44& 0.04& 0.02\\
        \hline
        Complete [ms] & 13207.24& 28.75& {\cellcolor[HTML]{C9C0BB}}{16.62}\\
        \hline
        Speedup & -& 459.46& {\cellcolor[HTML]{C9C0BB}}{1.73}\\
        Tot. speedup & -& 459.46& {\cellcolor[HTML]{C9C0BB}}{794.50}\\
        Memory [kB] & 920.6& 243& {\cellcolor[HTML]{C9C0BB}}{36.1}\\
        MOPS& 0.14& 231& {\cellcolor[HTML]{C9C0BB}}{268}\\\hline
    \end{tabular}
    }
    \caption{Performance and speedup analysis of EEGNet optimizations on {\pace} at 12.5MHz. The labels A–B correspond to the optimization steps detailed in section 3.3.}
    \label{tab:optimization_study}
\end{table}
The performance evaluation in Table~\ref{tab:optimization_study} quantifies the effect of hardware-aware optimizations applied to EEGNet on the {\pace} SoC’s CGRA. The baseline implementation on the Rocket core, lacking specialized neural acceleration, exhibited the highest latency and memory usage. Optimization A, which introduced 6-bit quantization and CGRA offloading, achieved a 460× speedup and reduced memory footprint from 920.6kB to 243kB. However, reliance on the Rocket core for ELU activation introduced latency overhead. Optimization B replaced ELU with ReLU, fused batch normalization, and removed data layout transpositions, enabling full offload to the CGRA and reducing inference time to 16.6ms. This corresponds to a total speedup of 794× and a memory footprint of just 36.1kB. \textbf{These results highlight the effectiveness of model-hardware co-design in dramatically reducing latency and memory requirements.}

\subsubsection{Hardware-Based Deep Learning Inference Comparison:}

\begin{figure}[htbp]
    \centering
    \begin{minipage}{\columnwidth}
        \centering
        \includegraphics[trim={3mm, 3mm, 3mm, 3mm}, clip, width=0.75\linewidth]{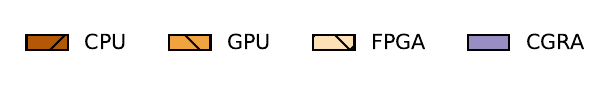}
        \vspace{-0.7em}
    \end{minipage}
    \begin{minipage}{\columnwidth}
    \begin{subfigure}[t]{0.49\textwidth}  
        \centering
        \includegraphics[trim={3.1mm, 4.7mm, 3mm, 3mm}, clip, width=\linewidth]{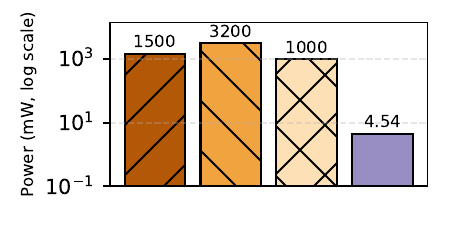}  
        \caption{Power comparison}
        \label{fig:power_comp}
    \end{subfigure}
    \begin{subfigure}[t]{0.49\textwidth}
        \centering
        \includegraphics[trim={3.1mm, 3.6mm, 3mm, 3mm}, clip, width=\linewidth]{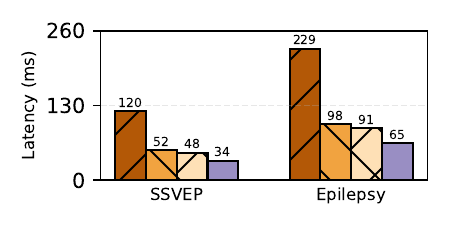}  
        \caption{Latency comparison}
        \label{fig:latency_comp}
    \end{subfigure}
     \end{minipage}
    \caption{Hardware performance comparison of LPEEGNet on CPU, GPU, FPGA, and CGRA ({\pace} SoC) in terms of power consumption and latency.}
    \label{fig:hardware_comparison}
\end{figure}

Fig.~\ref{fig:hardware_comparison} presents a comparative analysis of power consumption (constant across tasks) and inference latency for both SSVEP and epilepsy detection, evaluated using EEGNet across multiple hardware platforms: Nucleo STM32 CPU~\cite{st_nucleo_l476rg}, Jetson Orin GPU~\cite{nvidia_jetson_orin}, Basys 3 FPGA~\cite{basys3}, and the CGRA-enabled EdgeWizard SoC. The STM32, lacking specialized accelerators, consumed 1.5W and exhibited the highest latency (120ms). The Jetson Orin GPU reduced latency to 52ms but at the cost of 3.2W power draw. The Basys 3 FPGA based on a low-power Xilinx Artix-7 architecture \cite{fpga_for_eeg}, improved efficiency with 1.0W power and 48ms latency.

In contrast, the CGRA in EdgeWizard achieved 34ms latency for SSVEP and 48ms for epilepsy detection at just 4.54mW power, yielding 59.19 GOPS/W. This outperforms the CPU (0.5 GOPS/W), GPU (1.2 GOPS/W), and FPGA (2.8 GOPS/W). Signal preprocessing (notch and band-pass filtering) is also accelerated on the CGRA at 1.6× and 2.2× speedups over CPU execution, while classification accuracy remains comparable to full-precision baselines. 

{\pace} offers a streamlined programming workflow starting from standard PyTorch model, which is fed into the compiler to generate and schedule a dataflow graph. Common layers are already supported, including fully connected (52 lines) and convolution (33 lines). This makes deployment significantly simpler than writing CUDA for GPUs, OpenMP for multi-core CPUs, or managing pragmas for FPGAs. For operators not present in the existing library, users can specify new kernels in C code, with only a few hours of development and optimization needed, keeping the framework generalizable across different neural network architectures. Each layer is loaded onto the CGRA via interface functions, with non-linear operations such as Softmax and ReLU executing on the host CPU.

\subsection{End-to-End System Evaluation}
We evaluate end-to-end performance using the SSVEP BCI application. The SSVEP study characterises core system metrics (latency, power, classification accuracy) under controlled ambulatory conditions. The complementary algorithmic epilepsy results appear in Section~\ref{subsec:on_chip_performance}.

\label{sys_performance}
\begin{figure}[h]
    \centering
    \resizebox{\columnwidth}{!} {
    \includegraphics[trim={40mm 50mm 80mm 40mm}, clip, width=\linewidth]{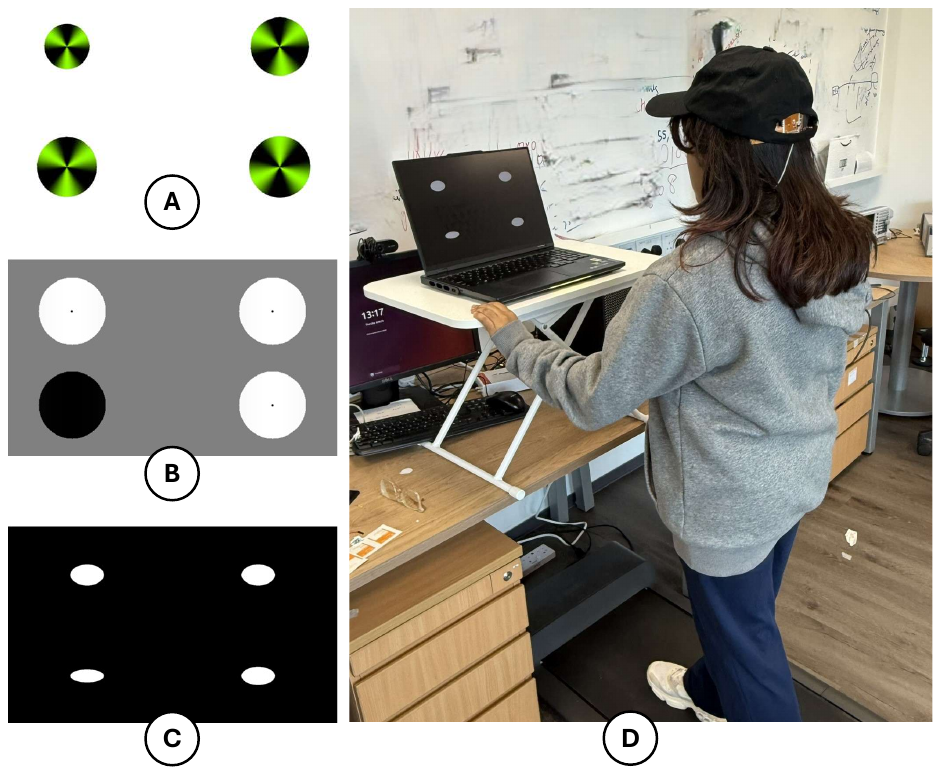}
    }
    \caption{(A) RadialZoom, (B) Flashing, and (C) Coin Flip SSVEP Stimulus. (D) Participant on treadmill for trial.}
    \label{fig:trial}
\end{figure}

\textbf{Methodology:}
We conducted a two-stage, Institutional Review Board (IRB)-approved user study using the SSVEP BCI paradigm to evaluate EdgeWizard’s end-to-end performance under both stationary and walking conditions. All metrics including classification accuracy, power consumption, inference latency were derived directly from these trials and benchmarked against state-of-the-art systems. In stage 1, fifteen healthy adults (25–40 years, normal or corrected vision) provided training data with our EEG headset. Participants experienced three SSVEP stimulus types: Radial Zoom \cite{radial_zoom}, Flashing \cite{WalkingWizard}, and Coin Flip \cite{coinflip} as shown in Fig.~\ref{fig:trial}(A,B,C). These stimuli were presented as four corner targets flickering at 6.0, 6.67, 7.5, and 8.57 Hz on a 60 Hz LED display. Participants fixated on each target for 40 s while standing and while walking at 3 km/h on a treadmill. Participants then rated each stimulus type for perceived visual fatigue.

In stage 2, the participants performed rapid‐classification trials using the stimulus identified by consensus as least fatigue - Coin Flip\cite{coinflip}. Each trial comprised 6.5 s fixation per target, two trials standing and two trials walking at 3.0km/hr, yielding test data for our deep-learning classifier. Fig.~\ref{fig:trial}D illustrates a participant during these sessions. We then deployed LPEEGNet on the EdgeWizard platform for real‐time decoding, and the resulting performance measurements are presented below.  

\subsubsection{Classification Accuracy and Ablation Study:}
\begin{figure}[h]
    \centering
    \resizebox{\columnwidth}{!} {
    \includegraphics[width=\linewidth]{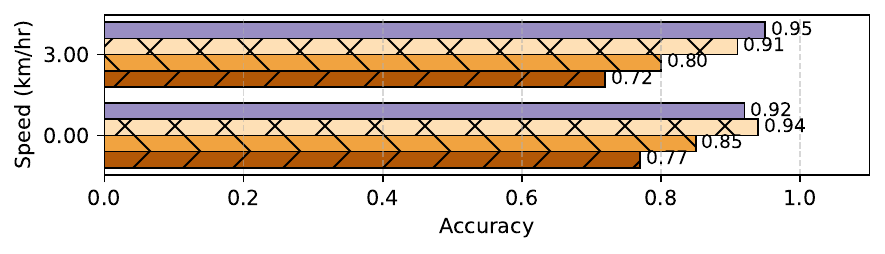}
    }
    \caption{Comparison of EdgeWizard SSVEP classification accuracy against state-of-the-art dry-electrode EEG. The accuracy performance for EdgeWizard as shown in \textcolor{lavender}{\rule{0.7em}{0.7em}} generally outperforms that of 
    \textcolor{burntOrange}{\rule{0.7em}{0.7em}}\cite{emotiv_7}, 
    \textcolor{softOrange}{\rule{0.7em}{0.7em}} \cite{mobileSSVEP_6} and
    \textcolor{lightPeach}{\rule{0.7em}{0.7em}}\cite{WalkingWizard}.}
    \label{fig:classification_accuracy}
\end{figure}
The EEG recordings of the Coin Flip SSVEP collected in stage 1 of the user trial were split into 80\% training and 20\% validation sets. Then, the EEG pre-processing steps described in Section~\ref{subsec:data_acq_model_development} were applied, with BPF setting from 5 to 30Hz, sliding window step-size of 0.2s, and a factor-4 downsample. LPEEGNet was then trained and loaded to {\pace} SoC. 

Fig.~\ref{fig:classification_accuracy} details EdgeWizard’s online classification performance during stage 2 of our user trial. The system achieved a mean online accuracy of 92\% under stationary conditions and 95\% during walking, \textbf{demonstrating robust and reliable decoding of user intent from streamed EEG across both static and mobile scenarios.} 

\textbf{Noise electrode contribution.} Ablation studies were conducted to assess the effect of the noise electrode. Removing the noise electrode reduces walking accuracy from 95\% to 83\% ($-$12~pp), while stationary accuracy is unchanged at 92\%, confirming that the noise electrode is critical specifically under motion, where it captures motion-induced artifacts that would otherwise corrupt the EEG channels.

In Section~\ref{subsec:motion_resilience}, we present results obtained when EdgeWizard was used during high-impact physical activities such as jogging, jumping, and simulated falls. In these scenarios, the EEG electrodes may momentarily lose contact with the scalp at the point of impact, during which the noise electrodes provide limited benefit. However, once contact between the EEG electrodes and the scalp is re-established and the user continues moving, the noise electrodes again become effective in capturing and mitigating motion-induced artifacts.

\textbf{Spring-loaded pin contribution.}
Direct isolation of the spring-loaded pin mechanism would require fabricating a separate PCB with fixed pins, which is beyond the scope of this work. We provide indirect evidence through two observations: (1) the impedance characterization in Fig.~\ref{fig:impedance} shows that the spring-loaded design achieves significantly lower electrode--scalp impedance than standard combed dry electrodes; and (2) the motion resilience measurements in Section~\ref{subsec:motion_resilience} show rapid signal recovery after high-motion events (Pearson $r=0.852$ post-jump). A fully isolated spring-pin ablation study remains future work.

\subsubsection{Inference power consumption and latency during online user trials:}
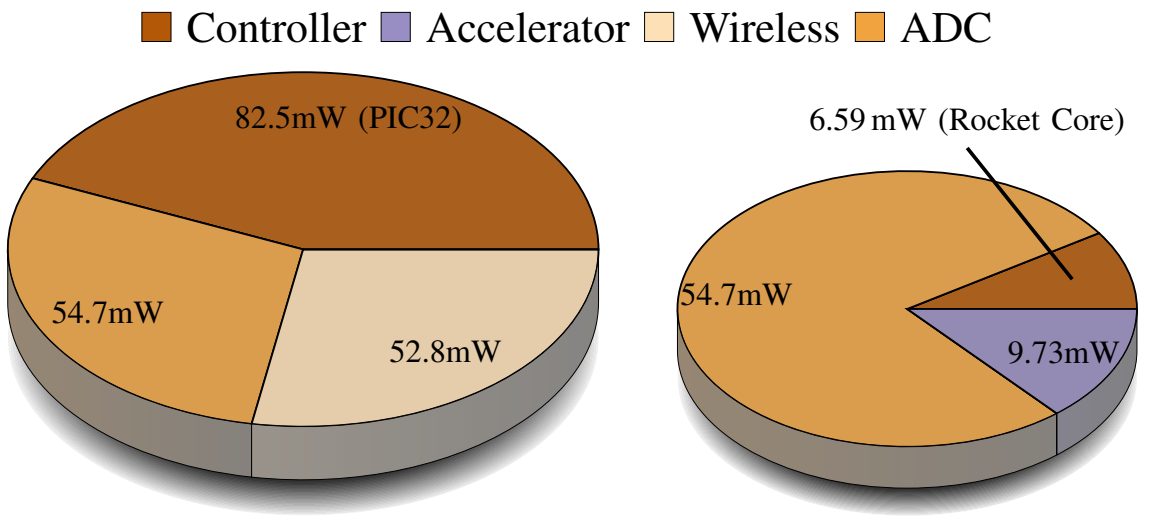
\begin{figure}[t] 
	\scriptsize
	\centering
    \captionsetup[subfloat]{labelfont=scriptsize,textfont=scriptsize}
    \resizebox{0.7\columnwidth}{!} {
\definecolor{customyellow}{HTML}{fbe4a2} 
\definecolor{customblue}{HTML}{cfeffa}
\begin{tikzpicture}
[
  acc/.style={shape=rectangle, draw=black,fill=lavender, line width=0.1},
  other/.style={shape=rectangle, draw=black,fill=manatee, line width=0.1},
  adc/.style={shape=rectangle, draw=black,fill=softOrange, line width=0.1},
  wireless/.style={shape=rectangle, draw=black,fill=lightPeach, line width=0.1},
  controller/.style={shape=rectangle, draw=black,fill=burntOrange, line width=0.1}]
  \matrix [node font=\scriptsize\large] at (-2,0) {
  \node [controller,label=right:Controller] {}; &
  \node [acc,label=right:Accelerator] {}; &
  \node [wireless,label=right:Wireless] {}; &
  \node [adc,label=right:ADC] {};\\
};
\end{tikzpicture}}
    \\
    \resizebox{0.9\columnwidth}{!} {
    \subfloat[\textbf{Off-board processing, 190 + P$_{ext}$ mW}]{
	\begin{tikzpicture}
\definecolor{customyellow}{HTML}{fbe4a2} 
\definecolor{customblue}{HTML}{cfeffa}
 \piechartthreed[scale=0.45,
                   background color=orange!50,
                   mix color= gray]
                   {156.312/burntOrange,0/manatee,103.644/softOrange,100.044/lightPeach,0/lavender}
\draw[black] (pc 1) coordinate (s1) node{82.5mW (PIC32)};
\draw[black] (pc 3) coordinate (s1) node{54.7mW};
\draw[black] (pc 4) coordinate (s1) node{52.8mW};
\end{tikzpicture}
    \label{fig:power_walkingwizard}
    }    
    \subfloat[\centering\textbf{Onboard processing \\ with EdgeWizard, 71.02mW}]{
	\begin{tikzpicture}
\definecolor{customyellow}{HTML}{fbe4a2} 
\definecolor{customblue}{HTML}{cfeffa}
 \piechartthreed[scale=0.35,
                   background color=orange!50,
                   mix color= gray]
                   {33.408/burntOrange,0/manatee,277.272/softOrange,0/lightPeach,49.32/lavender}

\coordinate (wedge1) at (pc 1);
\coordinate (label1) at ($(wedge1) + (-0.8, 1)$);  
\draw[-, thick] (wedge1) -- (label1) node[above]{6.59\,mW (Rocket Core)};

\draw[black] (pc 3) coordinate (s1) node{54.7mW};
\draw[black] (pc 5) coordinate (s1) node{9.73mW};
\end{tikzpicture}
    \label{fig:power_edgewizard}
    }
    }
\caption{Power consumption comparison between off-board processing and on-board processing with EdgeWizard. For the former, \(P_{ext}\) represents the power consumption of the external workstation (not included in the numbers)} 
	\label{fig:power_ablation}
\end{figure}

\begin{figure}[t] 
	\scriptsize
	\centering
    \captionsetup[subfloat]{labelfont=scriptsize,textfont=scriptsize}
    \resizebox{0.7\columnwidth}{!} {
    \begin{tikzpicture}
[
  inference/.style={shape=rectangle, draw=black,fill=lavender, line width=0.1},
  other/.style={shape=rectangle, draw=black,fill=manatee, line width=0.1},
  notch/.style={shape=rectangle, draw=black,fill=softOrange, line width=0.1},
  bpf/.style={shape=rectangle, draw=black,fill=lightPeach, line width=0.1},
  getsamples/.style={shape=rectangle, draw=black,fill=burntOrange, line width=0.1}]
  
  \matrix [node font=\scriptsize\large] at (-2,0) {
  \node [inference,label=right:Inference] {}; &
  \node [notch,label=right:Notch] {}; &
  \node [bpf,label=right:BPF] {}; &
  \node [getsamples,label=right:Get Samples] {};\\
};
\end{tikzpicture}

}
    \resizebox{0.9\columnwidth}{!}{
    \subfloat[\textbf{Off-board processing, 2031.65 ms}]{
	\begin{tikzpicture}
\definecolor{customyellow}{HTML}{fbe4a2} 
\definecolor{customblue}{HTML}{cfeffa}
 \piechartthreed[scale=0.45,
                   background color=orange!50,
                   mix color= gray]
                   {79.48/burntOrange,0/manatee, 1.7/softOrange, 3.19/lightPeach, 275.63/lavender}
\draw[black] (pc 1) coordinate (s1) node{448.54 ms};

\coordinate (wedge3) at (pc 3);
\coordinate (label3) at ($(wedge3) + (1.25, 0.25)$);  
\draw[-, thick] (wedge3) -- (label3) node[right]{9.6 ms};

\coordinate (wedge4) at (pc 4);
\coordinate (label4) at ($(wedge4) + (-1.5, 0.5)$);  
\draw[-, thick] (wedge4) -- (label4) node[left]{18 ms};

\draw[black] (pc 5) coordinate (10,-1.2) node{1555.51 ms};


\end{tikzpicture}
    \label{fig:latency_walkingwizard}
    }    
    \subfloat[\centering\textbf{Onboard processing \\ with EdgeWizard, 34.01 ms}]{
	\begin{tikzpicture}
\definecolor{customyellow}{HTML}{fbe4a2} 
\definecolor{customblue}{HTML}{cfeffa}
 \piechartthreed[scale=0.35,
                   background color=orange!50,
                   mix color= gray]
                   {169.362/burntOrange,0/manatee,5.293/softOrange,9.421/lightPeach,175.924/lavender}

\draw[black] (pc 1) coordinate (s1) node{16 ms};

\coordinate (wedge3) at (pc 3);
\coordinate (label3) at ($(wedge3) + (0.25, 1)$);  
\draw[-, thick] (wedge3) -- (label3) node[above]{0.5 ms};

\coordinate (wedge4) at (pc 4);
\coordinate (label4) at ($(wedge4) + (-0.25, 1.5)$);  
\coordinate (wedge4left) at ($(wedge4) + (-0.2, 0)$);
\draw[-, thick] (wedge4left) -- (label4) node[above]{0.89 ms};

\draw[black] (pc 5) coordinate (s1) node{16.62 ms};
\end{tikzpicture}
    \label{fig:latency_edgewizard}
    }
    }
    \caption{Latency comparison between system with off-board processing and EdgeWizard with on-board processing, evaluated at a 12.5MHz SoC clock. For the off-board processing setup, \textit{Get Samples} includes both signal acquisition and wireless transmission via Bluetooth.} 
	\label{fig:latency_ablation}
\end{figure}
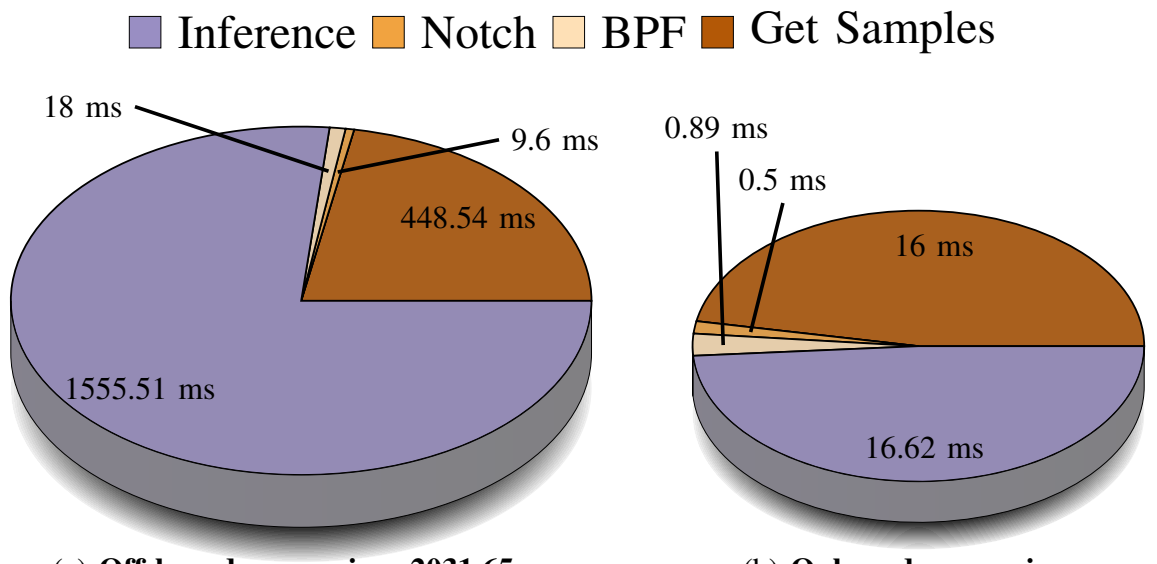

EdgeWizard’s fully integrated architecture consumes only 71.0mW during online operation, with 54.7mW from the ADS1299 analog frontend, 6.6mW from the RISC-V Rocket core, and 9.7mW from the CGRA accelerator as illustrated in Fig.\ref{fig:power_ablation}. For low-power EEG applications, replacing the ADS1299 with a custom Successive Approximation Register (SAR) ADC could reduce AFE power to the microwatt range, as shown in prior work \cite{sar_adc, sar_adc2}. In comparison, a traditional off-board system~\cite{WalkingWizard} uses a PIC32 microcontroller for wireless streaming (82.5mW), in addition to the ADC (54.7mW) and wireless module (52.8mW), totaling about 190mW on-head before considering external compute. This compute is ran on a typical workstation with an Intel Core i9-13900HX CPU (2.2GHz) and NVIDIA RTX 4090 GPU (2.5GHz), which incurs significant system power overhead. EdgeWizard can operate for up to 23 hours on a 500mAh battery, compared to under 9 hours for the off-board system, not accounting for the workstation’s power consumption.

For latency, EdgeWizard completes end-to-end SSVEP inference in 34.0ms at a 12.5MHz SoC clock (Fig.\ref{fig:latency_ablation}), including signal acquisition, preprocessing, and LPEEGNet inference on the Rocket core and CGRA. The off-board baseline requires 448.5ms for signal acquisition and Bluetooth transfer, plus 0.8ms notch filtering, 1.96ms bandpass filtering, and 51.6ms model inference - totaling 502.9ms. CGRA offloading of the filter stages further reduces their combined contribution to under 2~ms ($<$7\% of total), a 1.6--2.2$\times$ improvement over CPU execution.

Increasing the SoC clock to 100MHz would reduce EdgeWizard’s latency to \~4.3ms while maintaining its energy efficiency.
\begin{tcolorbox}[
  colback=boxbg,        
  colframe=black,       
  boxrule=1pt,          
  arc=2pt,
  left=2pt, right=2pt, top=2pt, bottom=2pt
]
\textbf{EdgeWizard achieves near-baseline accuracy with significantly lower latency and power, confirming its suitability for efficient, on-device EEG inference in wearable applications.}
\end{tcolorbox}

\subsection{Motion Resilience Across Physical Activities}
\label{subsec:motion_resilience}
\begin{figure}[h]
    \centering
    \includegraphics[trim={4mm 25mm 22mm 30mm}, clip, width=\linewidth]{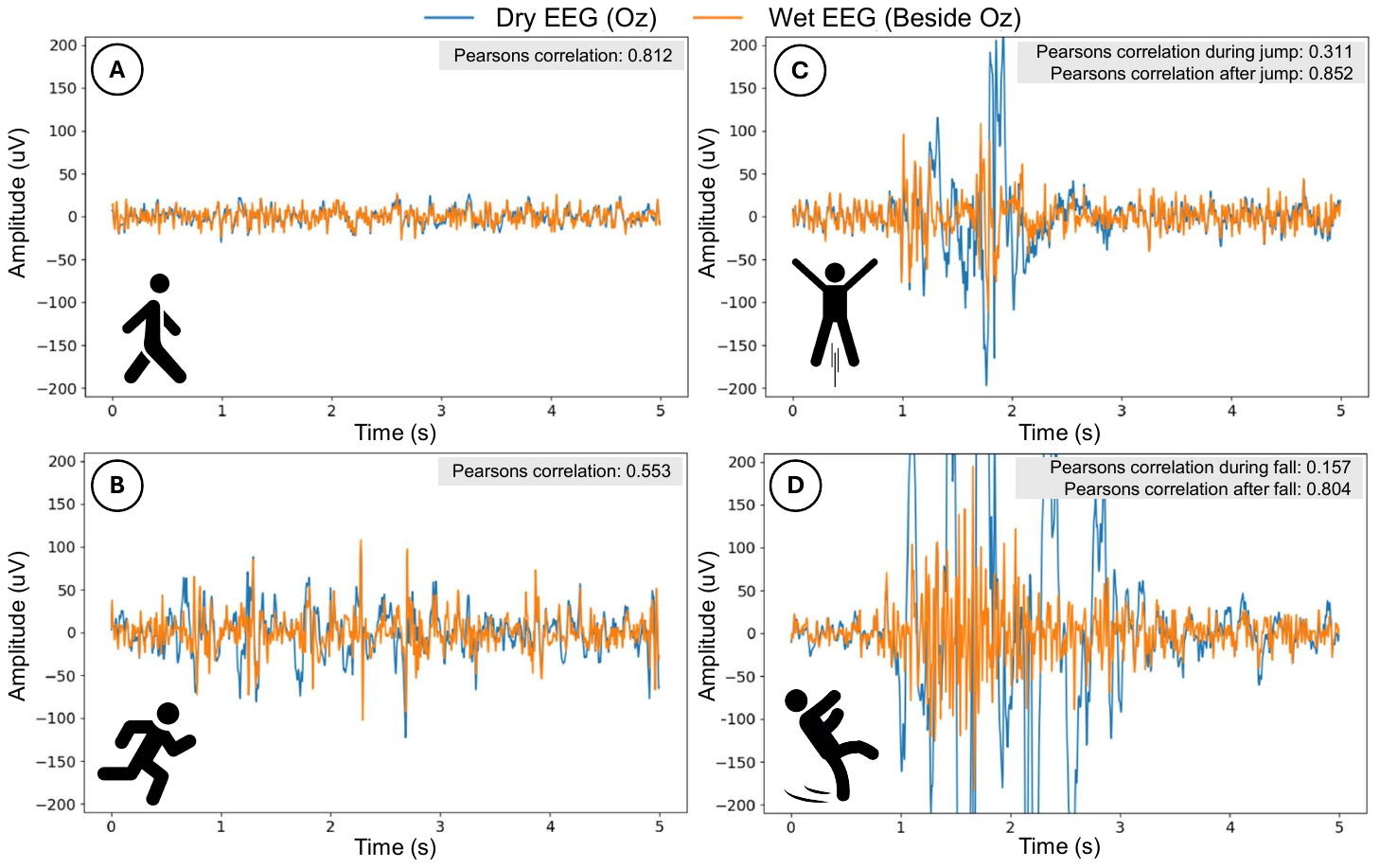}
    \caption{Evaluation of EdgeWizard under various physical activities: (A) Slow walking (up to 3.0km/hr), (B) Jogging, (C) Jumping on the spot, and (D) Simulated fall.}
    \label{fig:physical_activities}
\end{figure}

To assess motion resilience, we adopted \cite{corr_motion_artifact_dry_electrode}'s methodology, comparing dry electrode signals against a gold-standard wet electrode. One of the unused ADS1299 channels (accessible via the 10-pin auxiliary connector, Fig.~\ref{fig:hw}D), was connected to a gold-cup wet electrode placed adjacent to the dry Oz site, secured with 3M micropore tape and Ten20 conductive paste as ground-truth.

Recordings were collected across four activities: slow walking ($\leq$3km/h), jogging, jumping on the spot, and a simulated fall. Figure~\ref{fig:physical_activities} compares the time-domain waveforms acquired from both electrode types. During slow walking, dry and wet signals show strong correlation (Pearson $r=0.812$), indicating negligible motion artifacts. Correlation decreases during jogging ($r=0.553$), but the dry signal recovers rapidly after high-motion events such as jumping ($r=0.311$ during, $0.852$ after) and simulated falls ($r=0.157$ during, $0.804$ after).

\begin{tcolorbox}[
  colback=boxbg,        
  colframe=black,         
  boxrule=1pt,            
  arc=2pt,                
  left=2pt, right=2pt, top=2pt, bottom=2pt
]
\textbf{This demonstrates that EdgeWizard maintains robust EEG acquisition even under high-motion scenarios, establishing its viability as wearable neurotechnology.}
\end{tcolorbox}

\subsection{User Feedback, Long-Term Stability and Robustness In Adverse Conditions}

\newcommand{\sqboxs}{1.2ex}
\newcommand{\sqboxf}{0.6pt}
\newcommand{\sqbox}[1]{\textcolor{#1}{\rule{\sqboxs}{\sqboxs}}}
\newcommand{\sqboxEmpty}[1]{%
  \begingroup
  \setlength{\fboxrule}{\sqboxf}%
  \setlength{\fboxsep}{-\fboxrule}%
  \textcolor{#1}{\fbox{\rule{0pt}{\sqboxs}\rule{\sqboxs}{0pt}}}%
  \endgroup
}

\begin{table}[ht]
\centering
\resizebox{\columnwidth}{!}{
\begin{tabular}{|c|c|c|}
\hline
\textbf{Criterion} & \textbf{Definition} & \textbf{User Score}\\
\hline
Easy to Use         & Effort required to perform routine tasks           & \sqbox{black} \sqbox{black} \sqbox{black} \sqbox{black} \sqbox{black}    \\
Intuitive           & How well the system matches user expectations      & \sqbox{black} \sqbox{black} \sqbox{black} \sqbox{black} \sqbox{black} \\
Consistency         & Uniform behavior across similar operations         & \sqbox{black} \sqbox{black} \sqbox{black} \sqbox{black} \sqboxEmpty{black} \\
Learnability        & Speed at which new users become proficient         & \sqbox{black} \sqbox{black} \sqbox{black} \sqbox{black} \sqboxEmpty{black} \\
Willingness to Reuse& Likelihood of choosing the device again            & \sqbox{black} \sqbox{black} \sqbox{black} \sqbox{black} \sqbox{black} \\
\hline
\end{tabular}
}
\caption{User evaluations on five usability criteria (1–5 scale). Filled boxes indicate score distribution.}
\label{tab:usability}
\end{table}
\textbf{User Feedback.} We conducted a small‐scale System Usability Scale (SUS) assessment using a standard five-point Likert questionnaire, administered to the participants immediately after their first interaction with EdgeWizard. As shown in Table~\ref{tab:usability}, all criteria achieved a median score of 5/5; only \textit{Consistency} and \textit{Learnability} received a single score of 4 among otherwise perfect ratings. Participants unanimously praised the interface for its clarity and responsiveness, and indicated strong willingness to adopt EdgeWizard again. The dip in \textit{Learnability} reflects the modest initial effort required to master electrode placement and protocol steps.

\begin{figure}[h]
    \centering
    \includegraphics[trim={25mm 35mm 30mm 35mm}, clip, width=\columnwidth]{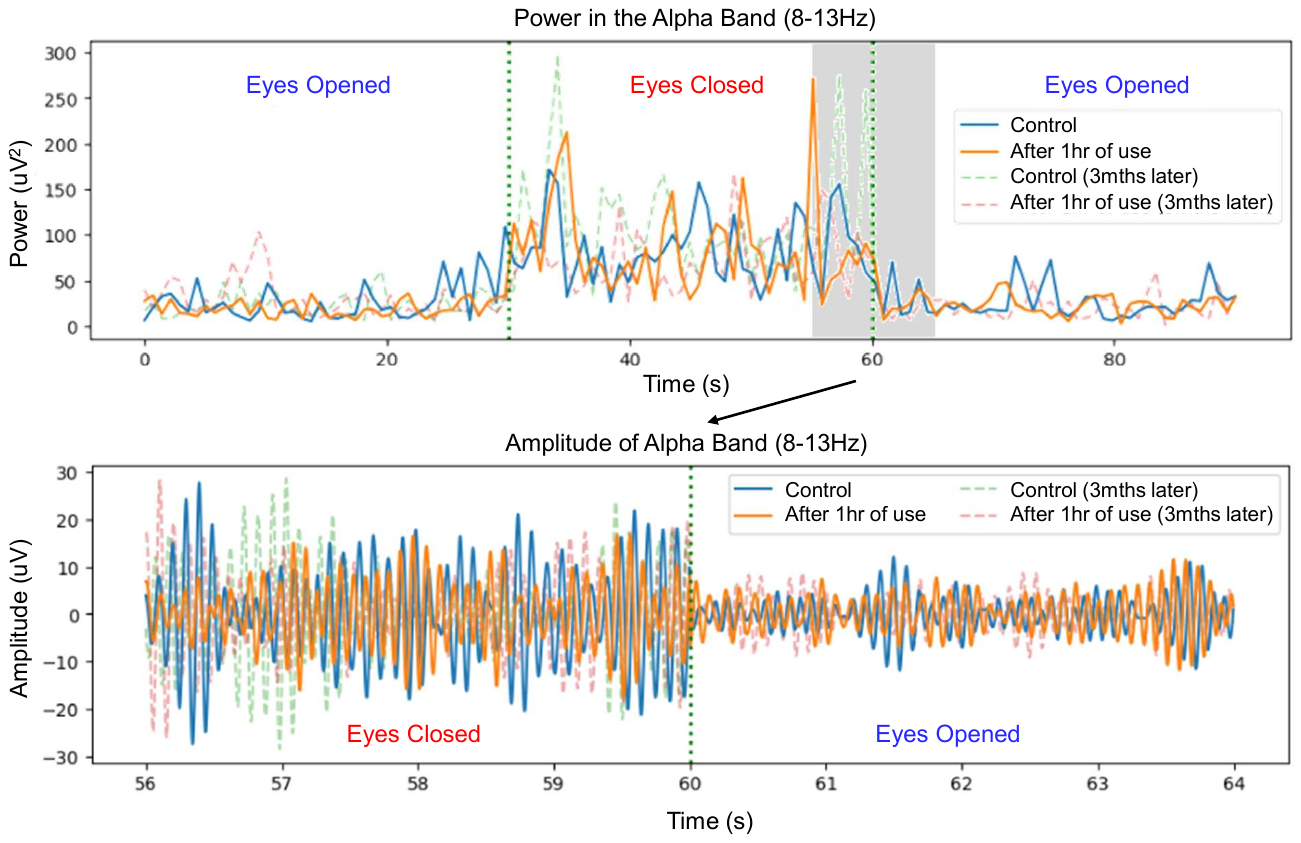}
    \caption{Assessment of long-term signal stability of EdgeWizard. Recordings were made after one hour of continuous use and repeated three months later. \textbf{Top:} Alpha band power (8–13Hz). \textbf{Bottom:} Raw EEG amplitude during transition from eyes closed to eyes open.}
    \label{fig:long_term}
\end{figure}

\textbf{Long Term Stability.} To evaluate the stability of the analog front-end, we recorded occipital EEG from a single subject under alternating eyes-closed and eyes-open conditions. The recording was performed immediately after power-up and again after one hour of continuous use. As shown in Fig.~\ref{fig:long_term}, we focused on the 8–13Hz alpha band and its well-known modulation pattern - elevated power with eyes closed and marked attenuation upon eye opening ("alpha blockade"). The top panel shows alpha power across time, while the bottom panel presents the corresponding amplitude–time waveform during the eyes-open transition. In both recordings, alpha power remained elevated with eyes closed and suppressed with eyes open, exhibiting no measurable drift in spectral power or amplitude after one hour. The same experiment was repeated three months later with consistent results, further validating the device’s long-term reliability.

\textbf{Robustness under Adverse Conditions.} To evaluate system performance under extreme physiological conditions, EdgeWizard was tested in a heat-stress scenario. The participant walked on a treadmill for 15 minutes at speeds up to 5.3km/h with an incline of up to 15°, inside a climate chamber maintained at 30°C and 70\% relative humidity. To mitigate excessive perspiration and prevent sweat accumulation near the electrodes, a thin layer of sodium polyacrylate superabsorbent polymer (SAP) was placed around the electrode sites to absorb moisture during the experiment.

EEG metrics were evaluated at the 5-minute and 15-minute marks using the Relative Gamma Power (RGP) and EEG Working Index (EWI). Consistent with prior studies~\cite{rgp_ewi}, both RGP and EWI increased under heat stress conditions. In contrast, the Fatigue Index (FI)~\cite{fatigue_index} showed minimal variation between the two measurements. Importantly, stable EEG recordings were maintained throughout the experiment despite elevated temperature, humidity, and perspiration, demonstrating the robustness of the EdgeWizard system under challenging environmental conditions.

\begin{tcolorbox}[
  colback=boxbg,        
  colframe=black,         
  boxrule=1pt,            
  arc=1pt,                
  left=1pt, right=1pt, top=1pt, bottom=1pt
]
\textbf{EdgeWizard demonstrates high usability, stable long-term signal quality, and resilience under physiologically and environmentally challenging conditions. These results highlight its suitability for extended neurophysiological monitoring and real-world BCI applications beyond laboratory settings}
\end{tcolorbox}
\section{Insights from End-to-End System-Hardware-Software design}
\label{sec5:discussion}
The development process of EdgeWizard provides several insights for motion-resilient on-device EEG systems. We tussled at the seams of different components and their constraints, which revealed many tradeoffs that shaped the final design.

\textbf{Sampling rate and compute timing constraints:} Compute must complete within the sliding window to maintain real-time operation, which places an upper bound on allowable inference latency and therefore requires the system clock to run sufficiently fast for the processing pipeline to finish within the window. At the same time, the ADC operates most reliably with an SPI clock of approximately 2.048MHz for data acquisition at 250SPS. Since the SPI clock is derived from the system clock through a hardware divider, the system clock cannot be arbitrarily scaled without affecting the achievable SPI frequency. As a result, both algorithmic latency requirements and SPI timing considerations constrain the system clock configuration.

{\bf System-level power tradeoffs:} At the algorithmic level, LPEEGNet trades model capacity for energy efficiency by constraining its memory footprint to the 512KB on-chip SRAM; while this avoids the power and latency overhead of off-chip SDRAM, it necessitates a highly compressed model architecture. The CGRA architecture reduces dynamic power via fine-grained clock gating of individual CGRA PEs during underutilized convolutional phases, which comes at the cost of control logic complexity to save power compared to fully-utilized dense layers.
SoC-level efficiency is maximized by power-gating the CPU during CGRA execution, which eliminates idle leakage at the cost of a marginal wake-up latency penalty incurred for inter-layer synchronization.
This optimizes the battery life by balancing performance of the model with aggressive power savings.

\textbf{PCB design considerations:} The original flex PCB only contained the ADC and electrode interface circuitry. Integrating the new SoC and supporting components significantly increased routing complexity, requiring a six-layer PCB to accommodate signal routing and power distribution. While this is atypical for flexible PCBs and introduces mechanical reliability concerns, we mitigated these issues through careful layout and mechanical reinforcement. Critical signal paths were routed with the shortest possible traces to reduce noise susceptibility, and mechanical stiffeners were added to support heavier components such as the main chip and memory. Since bending can introduce stress and signal integrity issues, the layout was designed so that most bending occurs in regions with minimal routing, reducing the likelihood of trace damage or electrical interference.

\section{Conclusion}
\label{sec6:conclusion}
EdgeWizard is the first fully motion-resistant, energy-efficient, real-time EEG wearable. It integrates LPEEGNet with a CGRA-based SoC for efficient on-device AI. By combining EEG signal preprocessing and deep learning inference, it achieves low-latency performance of 34ms with 23 hours of battery life on a 500mAh battery, ensuring reliable operation in dynamic environments. 

Designed for motion-tolerant BCI applications, EdgeWizard advances wearable neurotechnology, offering a scalable, high-performance platform for next-generation brain-computer interfaces.

\bibliographystyle{IEEEtran}
\bibliography{PACE-EEG-bib}

\begin{IEEEbiography}[{\includegraphics[width=1in,height=1.25in,clip,keepaspectratio]{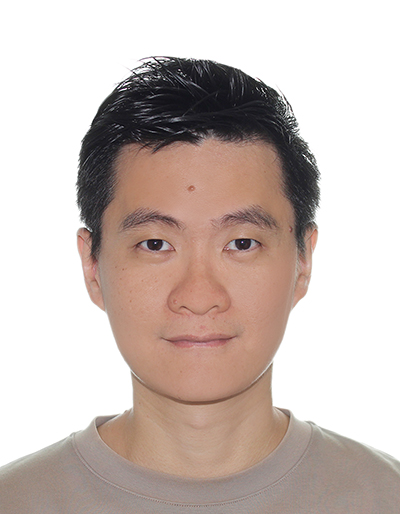}}]{Teck Lun} received the B.Eng. degree in Electrical and Electronic Engineering and the M.Sc. degree in Electronics from Nanyang Technological University, Singapore, in 2009 and 2010, respectively. Since 2010, he has been a Systems Engineer at DSO National Laboratories, Singapore, where he works on the development of advanced electronic systems. In 2025, he received the Ph.D. degree from the School of Computing, National University of Singapore, advised by Prof. Li-Shiuan Peh. His research interests lie at the intersection of system design and artificial intelligence, with a focus on EEG-based brain–computer interfaces.
\end{IEEEbiography}

\begin{IEEEbiography}[{\includegraphics[width=1in,height=1.25in,clip,keepaspectratio]{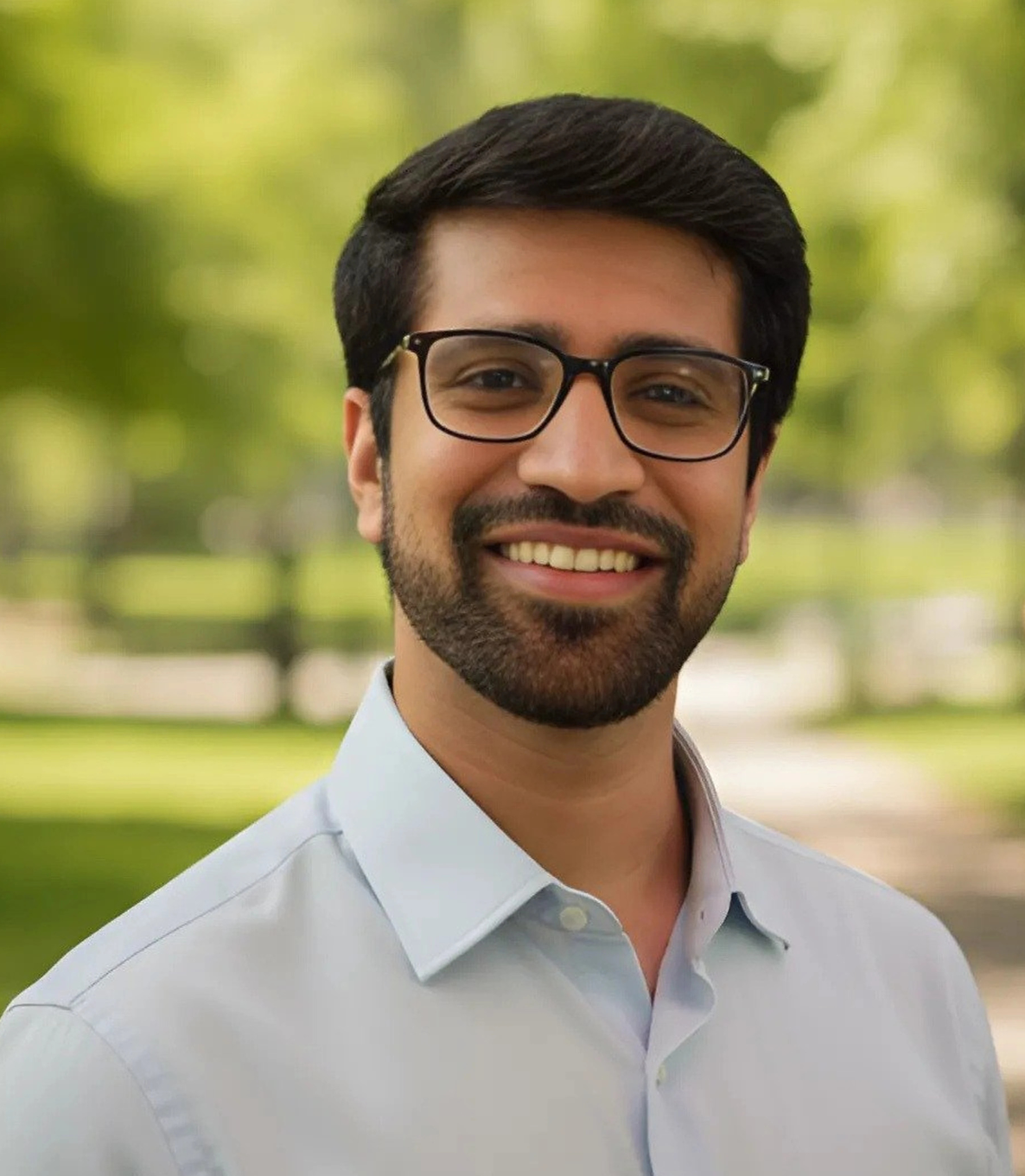}}]{Rohan Juneja} received the B.Tech. degree in Electronics and Communication engineering from IIIT-Delhi in 2018. In 2026, he received his Ph.D. degree from the School of Computing, National University of Singapore (NUS), Singapore, where he was co-advised by Prof. Li-Shiuan Peh and Prof. Tulika Mitra. From 2018 to 2021, he was with Qualcomm, where he contributed to the development of CPUs for Snapdragon chipsets. His research interests sit at the intersection of computer architecture, VLSI, and artificial intelligence, with a specific focus on building reconfigurable fabrics, and developing hardware-aware techniques for quantized and sparse models. He has published his research in several premier venues, including MICRO, ASPLOS, AAAI, ICCAD, DAC, HotChips.
\end{IEEEbiography}

\begin{IEEEbiography}[{\includegraphics[width=1in,height=1.25in,clip,keepaspectratio]{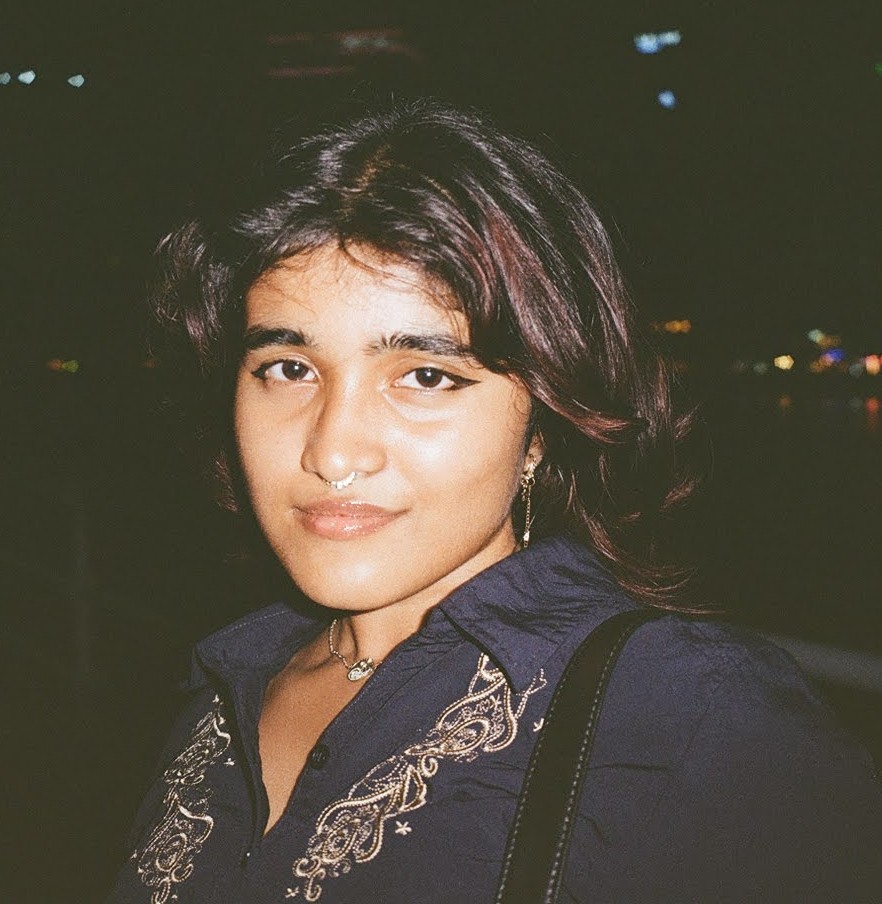}}]{Vishruti Ranjan} is a PhD student in Computer Science at the National University of Singapore (NUS), supervised by Prof. Li-Shiuan Peh since 2024. She received her B.Eng. (Hons) in Computer Engineering with Highest Distinction from NUS in 2023. Her research focuses on on-body neural sensing and feedback systems, sometimes coupled with reconfigurable computing for real-time applications. Her work is interdisciplinary, drawing insights from cognitive science, neuroscience, and human–computer interaction to design efficient, low-power systems.
\end{IEEEbiography}

\begin{IEEEbiography}[{\includegraphics[width=1in,height=1.25in,clip,keepaspectratio]{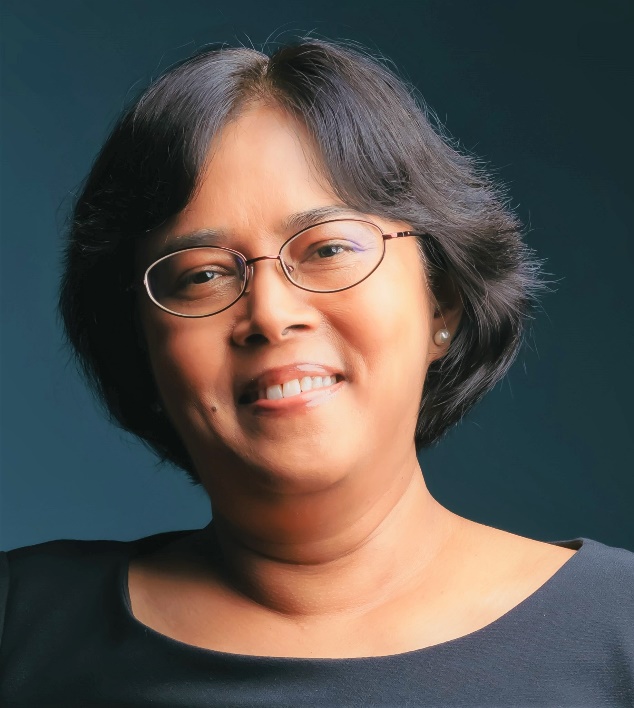}}]{Tulika Mitra} (Member, IEEE) received the B.E. degree in computer science from Jadavpur University, Kolkata, India, in 1995, the M.E. degree in computer science from the Indian Institute of Science, Bengaluru, India, in 1997, and the Ph.D. degree from the State University of New York, Stony Brook, NY, USA, in 2000. She is currently Dean and Provost’s Chair Professor of computer science with the School of Computing, National University of Singapore, Singapore. Her research interests include the design automation of embedded realtime systems with particular emphasis on software timing analysis/optimizations, application-specific processors, energy-efficient computing, and heterogeneous computing.
\end{IEEEbiography}

\begin{IEEEbiography}[{\includegraphics[width=1in,height=1.25in,clip,keepaspectratio]{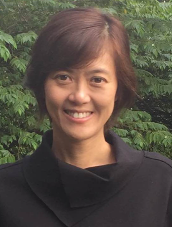}}]{Li-Shiuan Peh} joined National University of Singapore (NUS) as Provost’s Chair Professor in the Department of Computer Science, with a courtesy appointment in the Department of Electrical and Computer Engineering in 2016. Previously, she was Professor of Electrical Engineering and Computer Science at MIT and was on the faculty of MIT since 2009. Prior to MIT, she was on the faculty of Princeton University from 2002. She graduated with a Ph.D. in Computer Science from Stanford University in 2001, and a B.S. in Computer Science from NUS in 1995. Her awards include 3 test-of-time paper awards (ACM SIGMOBILE 2017 for ASPLOS 2002 paper, IEEE TCCA 2020 for HPCA 2001 paper, ACM SIGMICRO 2023 for MICRO 2002 paper), IEEE Fellow (2017), NRF Returning Singaporean Scientist Award (2016), ACM Distinguished Scientist Award (2011), MICRO Hall of Fame (2011), CRA Anita Borg Early Career Award (2007), Sloan Research Fellowship (2006), and the NSF CAREER award (2003).
\end{IEEEbiography}

\vfill

\end{document}